\documentclass[aps,prd,superscriptaddress,eqsecnum,amsfonts,showpacs,epsfig]{revtex4}
\usepackage{graphicx}

\newcommand{\be}{\begin{equation}}
\newcommand{\ee}{\end{equation}}
\newcommand{\bea}{\begin{eqnarray}}
\newcommand{\eea}{\end{eqnarray}}

\newcommand{\nn}{\nonumber}
\newcommand{\ep}{i\epsilon}
\newcommand{\om}{\omega}

\begin{document}

\title{Muon pair production with hadronic vacuum polarization re-evaluated
using precise data fits}
\author{V. \v{S}auli  }
\affiliation{Department of Theoretical Physics, Institute of Nuclear Physics Rez near Prague, CAS, Czech Republic  }

\begin{abstract}
Using optical theorem within sophisticated fits for exclusive hadronic productions in $e^+e^-$ collisions
the interference between leptonic and hadronic vacuum polarization functions is considered
 and applied for calculation of $\mu-$pair productions below 3 GeV. 
 Particularly, the dispersion relation for the  pion production form factor was obtained, confirming 
 a possible evidence for the light
  $\rho(1250)$  resonance. The  inflated error was introduced 
 and used to discriminate the quality of various  phenomenological fits.
 Resulting hadronic vacuum polarization was compared with  other conventional sources
 and  compared with KLOE experiment for $\mu^{-}\mu^{+}$ production at $\phi $ meson energy.
Furthermore, the obtained running fine structure coupling is  compared with  KLOE2 experiment for radiative return  
$\mu^{-}\mu^{+}$ production at $ \omega/\rho $ meson energy and careful analysis of the error stemming form various fits 
of combined data  was made and discussed.
  
\end{abstract}


\maketitle

%%%%%%%%%%%%%%%%%%%%%%%%%%%%%%%%%%%%%%%%%%%%%%%%%%%%%%%%%%%%%%%%%%%%%%%%%%%%%%%%%%%%%

%
\section{Introduction}

Comparisons between theory and experiment are used to test Standard Theory for decades. 
  For an accurate measurement the studies
require consistent account of leptonic as well as hadronic virtual corrections. The hadronic contribution to photon  vacuum 
polarization function plays particularly important role, since it is the main source of uncertainties in theoretical calculation
of muon anomalous magnetic moment $a_{\mu}$. The last precise measurement of $a_{\mu}$, together with the last decades data 
for electrohadron production, confirms the evidence of tension between Standard Theory and experiments  \cite{amu1,amu2,Abi2021}. 
Similar confrontation of the theoretical technique with the experimental accuracy is offered by  observed \cite{AUGU1973,BERKOM1976} interference effect between leptonic and hadronic vacuum polarization functions at energies of   narrow resonances: $\omega$ and $\phi$ and  heavier quarkonia $J/\psi,\Psi$ and  $\Upsilon$'s. There, the total cross section $\sigma_h$ is  enhanced several orders of magnitude when compared to other region of energy.
In practice, the effect is explored in the so called B-factories like BABAR, BELLE or BESS or more earlier
in Frascati where detectors and accelerator are tuned for $\Upsilon$ or $\phi$ meson energy. Narrow mesons  
 off prints in the vacuum polarization  turns to be  a measurable  fluctuation in the QED running coupling- the electromagnetic decay of mesons. This effect is presented in the all electromagnetic processes, being most easily observed in a single  Mandelstam variable  $s$-dependent process, e.g. the muon pair production. 
The most recent precise measurements of muon production $e^{+}e^{-}\to \mu^{+}\mu^{-}$ by  SND, CMD-2,3 and the KLOE(2) detectors represent another possible stringent    test of the Standard  Theory. In this case,   the theory and experiment comparison means to compare the whole shape of energy dependent functions.
 
With incomes of many new precise data for the cross section $\sigma_h$ we reanalyzed the 
calculation for \cite{AMBRO2005} within the use polarization function obtained by groups \cite{polar2,polar3} as well as by independent calculation performed here.
Further, having extracted hadron polarization function, we provide independent  comparison of  theory and recent KLOE2 \cite{KLOEdva} measurement of the fine structure coupling constant. 

Particularly, we performed the fit for  amplitudes of all necessary hadronic cross sections with emphasizing on two pions channel, 
for which we compare two rather different approaches. 
The $\pi\pi$ production cross section represents the most dominant hadronic  contribution ( 75 $\%$) to the muon anomalous moment and  
it should contain  important information about spectra of $1^-$ vector meson resonances. Notably, albeit only for spacelike momentum $Q^2$, it involves 
 the form factor function, which has been calculated from Schwinger-Dyson equations (the nonperturbative set of equation for Green's functions).
  At opposite -timelike-  momentum axis, the position and  the ordering  of resonances is believed to be important for understanding of confinement 
issue. Let us remain the old-fashionable quantum mechanical picture  of  linear Regge trajectories $M^2(N)$ where $N$ is radial number 
of the resonance. In this respect a recent evidence for very wide resonance $\rho(1250)$  from the unitary multichannel reanalysis \cite{rupp}
 of elastic pion scattering  is in a certain  contradiction with the  modern Schwinger-Dyson equations studies (see for instance  \cite{HGK2017}).
For this purpose  we write down the dispersion relation for the pion form factor 
and fit the spectral function.  Reduction of number of parameters needed for such a description 
 shows the right analytical form has been chosen in this case. As a matter of the fact, making a precise fit of  combined data  form several most important experiments, 
 our results indicate  a broad vector meson like structure in the range $1.15-1.3 GeV$.
  It is also shown that non-resonant background plays an important role in the description.

\section{$\sigma_{\mu\mu}$ for KLOE 2004}

The theory and the comparison between calculated  cross section 
$\sigma_{\mu\mu}=\sigma(e^{+}e^{-}\to \mu^{+}\mu^{-}) $ and the high precision measurement obtained by KLOE detector \cite{AMBRO2005} is presented in this section. 

The integrated cross section, for which we adopt approximation and conventions given by \cite{ARBU1997}, can be written in the following way
\be \label{etomu}
\sigma(s)=\frac{4\pi C_t}{\left|1-\Pi(s)\right|^2}
\left[\sigma_A(s)\left(2-\beta_{\mu}^2(1-\frac{C_t^2}{3})\right)+\sigma_B(s)\right]
 \, ,
\ee
where $C_t$ stand for $\cos(\theta_{min})$ with $ \theta_{min}=50^0$ ($\theta_{max}=140^o$), which is KLOE  experimental cut on polar scattering angle between $\mu^{-}$ and $e^{-}$ particles and $\beta_{\mu}=\sqrt{1-4m^2_{\mu}/s}$.
The function $\sigma_A(s)$  is defined such that it  has an angular dependence   identical to the Born cross section. The rest is unique  and explicitly reads   
\be
\sigma_B(s)=-\frac{\alpha^3}{4\pi s}
(1-\beta_{\mu}^2)\ln{\frac{1+\beta_{\mu}}{1-\beta_{\mu}}} \, .
\ee
Thus the main term  $\sigma_A(s)$, listed completely in \cite{ARBU1997}, collects all  leading logs of Dirac and Pauli form factors and the known soft photon contributions for which we take $ln\frac{\Delta \epsilon}{\epsilon}=0.05$ (15 MeV cut on c.m.s. soft photon energy at $\phi$ peak).  
Let us mention that the the both $\sigma_A$ as well as $\sigma_B$ are slowly changing  real valued functions and do not play an important role in the observed interference effect.

The integral cross section formula is proportional to the square of the fine structure constant $\alpha(s)$, which reads
\be \label{alfa}
\alpha(s)=\frac{\alpha}{1-\Pi(s)} \, ,
\ee
with $\alpha=\alpha(0)=1/137.0359991390$ and where the polarization function
$\Pi(s)=\Pi_l(s)+\Pi_h(s) $ is completed from the leptonic $l$ and the  hadronic $h$ part.

Purely QED contributions are  well known from perturbation theory.
Since, there are some mistakes in the formula  in Ref. \cite{ARBU1997}, I present leptonic contribution into the vacuum polarization function here:
\be
\Pi_l(s)=\frac{\alpha}{\pi}\Pi_1(s)+\left(\frac{\alpha}{\pi}\right)^2\Pi_{2e}(s) \, \, ,
\ee
where one loop contribution is
\bea \label{lepton}
\Pi_1(s)&=&\Pi_e(s)+\Pi_{\mu}(s)+\Pi_{\tau}(s)\, \, ;
\nn \\
\Pi_f(s)&=&-5/9-x_f/3+f(x_f)\,\, ; f=e,\mu,\tau \, ; 
\nn \\
f(x_f)&=&\frac{\beta_f}{6}(2+x_f) 
\left(\ln{\frac{1+\beta_{f}}{1-\beta_f}} -i\pi\right)\Theta(1-x_f)
\nn \\
&+&\frac{\beta_f}{3}(2+x_f)  arctg\left({\frac{1}{\beta_f}}\right)\Theta(x_f-1)  \, \, ,
\eea
where $\beta_f=\sqrt{|1-x_f|}$ and $x_f=4m^2_f/s$.
Also the leading logarithmic term:
\be
\Pi_{2e}(s)=\frac{1}{4}ln(\frac{s}{m_e^2}-i\pi)+\zeta(3)-5/24 \, ,
\ee
which  stems from the second order is taken into account (for heavy quarks and large $q^2$ one can employ perturbation theory as well, remind only the usual extra factor $\alpha\rightarrow \alpha N_c e_q^2$ in the appropriate one loop expression).

Hadronic part of the polarization function  $\Pi_h$ is in principle  directly calculable  from the equations of motions \cite{GFW2011,sauli2020}
where  the later approach provides the first, albeit partially incomplete result in the  entire Minkowski space.
In approaches \cite{GFW2011,sauli2020}, HVP is calculated from the Standard Model parameters,i.e.  without a large use of experimental data.
Due to required high accuracy,  here we follow the old-fashionable approach and by using the Unitarity, 
the function  $\Pi_h$ is  extracted from the total hadronic production $\sigma_h=\sigma_{tot}(e^+e^-\rightarrow hadrons) $
and HVP is obtained  from the following dispersion relation \cite{CABGAT1961,EIDJEg1995}:
\be \label{muf}
\Pi_h(s)=\frac{s}{4\pi^2\alpha}\int_{m_{\pi}^2}^{\infty}d\omega \frac{\sigma_h(\omega) \left[\frac{\alpha}{\alpha(\omega)}\right]^2}
{\omega-s+i\epsilon}\, .
\ee

 Since the measurement of low energy inclusive cross section $\sigma_h$ is technically demanding,
 the knowledge of $\sigma_h$ relies on  many  experimental measurements of the {\it hadronic exclusive (hex)} processes $\sigma_{hex}$,
which constitute $\sigma_h$
\be
\sigma_h=\sum_{hex}\sigma_{hex} \, \, ,
\ee
 noting the   photons emitted from  the final hadronic states counts as well.

The expression (\ref{muf}) represents a nonlinear integral equation  with the singular kernel and to this point,
 there are only few groups \cite{polar2,polar3,DAVIER2011}, which steadily  collect necessary data on $\sigma_h$   and provide
  fresh and more accurate  look at the $\Pi_h$ line-shape.  
Before providing the result we describe the determination of $\sigma_h$ in our case.

First of all, we follow a common practice and very narrow resonances like  heavy quarkonia  replace by their  Breit-Wigner functions with PDG parameters. 
The inclusion of the entire rest of $\sigma_h$ is included numerically and represent the core of the method.
Since the kernel in Eq. (\ref{muf}) is singular
a straight use of experimental data would lead to uncontrolled numerical noise and lost of accuracy.
In order to  make the integration under control we construct the analytic fit of $\sigma_h$ as a first step.
We established the inflated error in the next section with main concern on the  the exclusive cross section $\sigma_{\pi^+\pi^-}$
  for which we construct our own fit based on spectral representation.

\section{ Fit for $\sigma_h$ and the inflated error}

Mutual incompatibilities in data originating from various experiment usually appear when combining and fitting the data from several different experiments. 
 They originate  in different systematical data errors, being always presented due to the different  experimental set up and as a consequence the standard fit criterion   $\chi^2\simeq 1$ turns to be problematic to fullfill for combined data.

Remind the reader, there are basically two methods to determine experimental hadronic cross section $\sigma_{h}$, the first method
scan energy intervals and  measure the hadronic cross section directly. 
The second one uses radiative return, being also named Initial State Radiation (ISR) method, wherein the $n$-body  hadronic cross section is  
extracted from $n+1$-body final hadron+photon  cross section.
The advantage of the later is minimized background and the access to an amazingly  large range of $s$ in single experiment.

 To be able to calculate the propagation of experimental errors, a successful introduction of semi-stochastic  inflated error (IE) is needed.
  Suitably defined inflated error $\sigma_I$ then can  partially accounts  (even not  well determined) systematic error in a statistical manner.
Since there does not exist a unique definition in the literature, we are going to define the IE through the  fit  $\sigma_{fit}$  of data $\sigma_h$ :  the IE $\sigma_I$ 
is defined such  that we get  minimal $\chi_I^2=1$ in a theoretically allowed space of functions $\sigma_{fit} $ and $\sigma_I$ and simultaneously $\sigma_I$ 
is smallest  for all  variables (energy $E_i$ in our case).
 More concretely, $\chi^2$ is  obtained through minimization of the following quantity
%$
\be \label{chicko}
\chi_I^2=\sum_i \frac{[\sigma_{h}(E_i)-\sigma_{fit}(E_i)]^2}{\sigma_I^2(E_i)}^2 \, .
\ee
where $E_i$ is the energy of positron-electron pair and $i$ runs over all data points.

Obviously,  the introduced  IE replaces the statistical  error $\sigma_{stat}$ of single experiment and  therefore 
as the additional requirement we impose the condition
 \be \label{ineq}
 \sigma_{stat}<\sigma_{I}
\ee 
 at all  points $E_i$. Such a construction automatically ensures that we must  get an upper estimate of 
 statistical error when single experiment is considered. 
 In general case, one should expect that  $\sigma_I \simeq \sigma_{tot}=\sqrt{\sigma_{stat.}^2+\sigma_{syst}^2}$ for many experiments with comparable accuracy.
 In this respect the  systematical errors are regarded as the additional noise.  
Of course, in case of small number of experiments, or in case of correlated experiments,
 the effect of systematical error should not be ignored and should be discussed separately.

 In suggested approach here, the IE is established simultaneously with the cross section fit  $\sigma_{fit}$ for each set of measured hadronic exclusive data. 
 More concretely we take
\be \label{maj}
\sigma_{I}(s)=c_I \sqrt{\sigma_{fit}(s)}
\ee
with a constant energy independent coefficient $c_I\simeq 1 nb^{1/2}$. It  provides all required properties we need for modern  data on $\sigma_h$ bellow $3GeV$.
We are satisfied with a single definition of the inflated error, however reaching a higher energies a split to  several definitions of IEs, each  valid only for a certain 
 interval could be done  in order to reflect accuracy at regions dominated by quarkonia.

Due to the simple choice (\ref{maj}), the value $c_I$ is therefore one of the main, but not unique, criterion of the quality of the data fir bellow $3GeV$.  
Depending on required or assumed  fit properties  (analyticity, spectrality, unitarity etc) there can exist more best fits due to other possible constrains.
Therefore  associated fits can have their own and in-equivalent IEs, nevertheless 
one can   compare them quantitatively by  achieved constants $c_I$. Alternatively one can evaluate the fit of  ``theory 1'' with the error achieved 
by another fit and compare their $\chi^2$ mutually.  Of course, the experimental statistical error can serve for comparison of various fits as well, albeit the associated error 
is underestimated, since counts only with individual statistical errors of principally different experiments.  
For  purpose of labeling we will use $\chi_D$ in order to distinguish from other $\chi^2$.

\section{$\pi\pi$ contribution to $\sigma_h$ via analytical pion form factor $F$ and evidence of $\rho(1250)$ }

\begin{figure}[htb]
\centerline{\includegraphics[width=7.5cm]{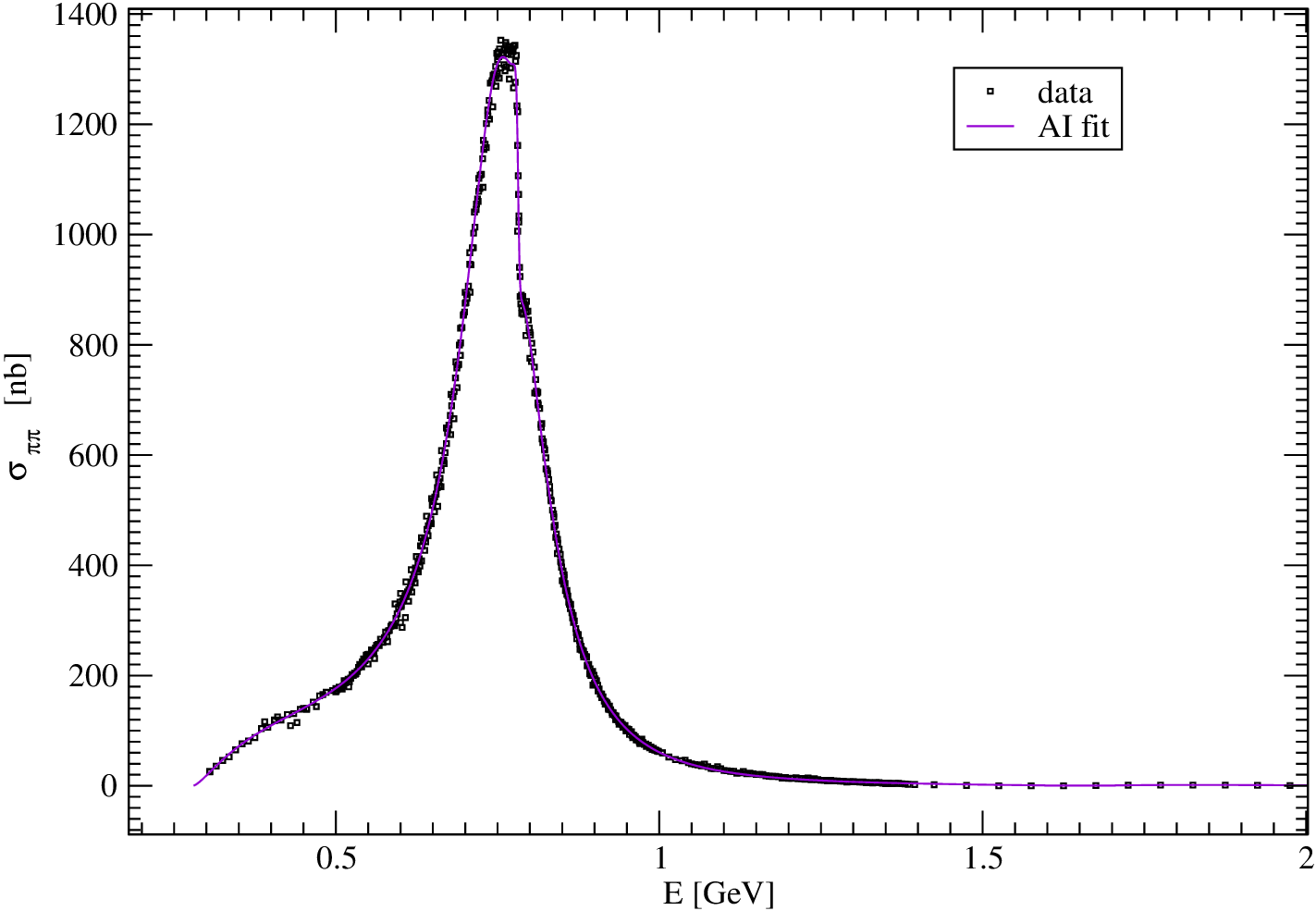} {\mbox , } \includegraphics[width=7.5cm]{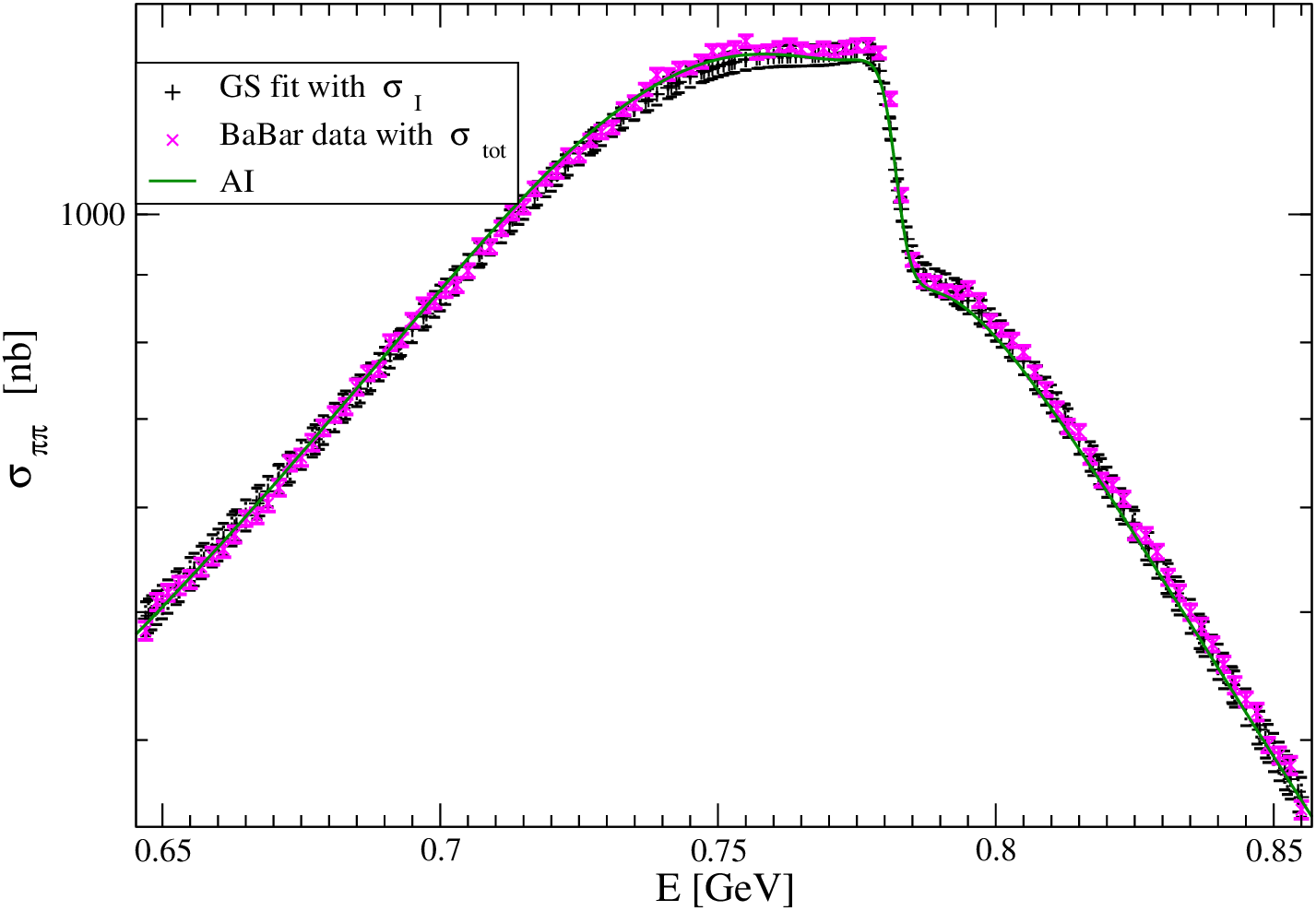}}
\caption{\label{nufik}$\sigma_{\pi\pi}$ cross section Left: All combined  data and the best spectral fit as described in text. Right: Here we show fits at $\rho$ peak position. GS fit evaluated at all data points  with inflated error is shown (light lines for errors). For comparison the  BaBaR data  with the total error are shown. The best spectral fit is shown as well (single black solid line). Errors are not shown for the later case in order  to keep other points visible.}
\label{figure5}
\end{figure}

We describe the fit for $\pi\pi$ production cross section in this Section. 
Likewise in  method used in \cite{rupp,BYKANA2016,BOLD2017} we impose a constrain and write down the fit for the amplitude.
For this purpose the spectral representation is chosen  for  the pion form factor, 
 which also involves meson $\rho(1250)$ with  three other  more conventional excited states of vector mesons. 
Imposing further constrains, which radically reduce the number of free parameters, still the excellent agreement with the data was found.
It provides the inflated error  determined by the constant $c_I^2=0.36 nb$, being thus substantial smaller then the result 
achieved previously in earlier analyses \cite{SAHVP1,SAHVP2}, where we got $c_I^2=0.64 nb$.

 The pion electromagnetic  form factor $ F_{\pi}$ is conventionally defined through  measured  hadronic cross section
\be
\sigma_{\pi\pi}(q)=\frac{\alpha^2\pi\beta^3}{3s} F_{\pi}^2(q) \, ,
\ee
i.e. with the vacuum polarization included in the form factor $F$.

Here we assume 
that  the entire pion form factor has  a real branch point corresponding to the two pionic threshold and therefore can be written 
in the form of the dispersion relation    
\be   \label{pionspec}
F_{\pi}(q)=\frac{1}{\pi}\int_{4 m_{\pi}^2}^{\infty} \frac{\rho_F(a)}{q^2-a+\ep}\, .
\ee

Thus this is the absorptive part $\Im F=-\rho_F$ which only needs to be known to determine $F$ entirely.
There  is  no any deep theoretical reason for the  analytic behavior dictated be single  variable dispersion (\ref{pionspec}) in QCD,
 however a similar dispersion relation are known to be  working in QCD/QED processes and provide a good agreement with 
 experiments even for more complicated kinematics  \cite{HOST2019}.

We do not calculate $F$ as we are unawared about working theoretical model which can provide 
correct result for energy above 1 GeV,  but we make a simple fit, which allows us to identify 
 vector resonances involved in the process. The heart of the fitting function for the pion charged form factor 
is the vector dominance model with sort of dressed Breit -Wigner peaks representing 
the resonances, which within  non-resonant background function $D_{bg}$ completes the result.

We look for  spectral function $\rho_F$ , which is searched in the following form
\be \label{mainfit}
\rho_F(a)= \Im \left\{{\cal D }_{bg}(a) \left[{\cal W}^{GS}_{\rho}(a,m_{\rho},\Gamma_{\rho})(s)\frac{1+c_{\om}{\cal W}_{\om}(a,m_{\om})}{1+c_{\om}}
+\sum_{i}c_{\rho^{i}}{\cal W}^{GS}_{\rho^i}(a,m_{\rho^i},\Gamma_{\rho^i})\right]\right\}\, \, .
\ee
where $c$ are real coefficients and  two parametric  function ${\cal W}^{GS}_{\rho^i}$ is listed in the Appendix.
The interference with the following ``background'' function 
\bea 
{\cal D}_{bg}(s)&=&{\cal N}e^{i\phi(s)}
\nn \\
 \label{bgfaze}
\phi(s)&=&\phi_0+\frac{1}{3}(1-4m^2_{\pi}/s)^{1/2} \, ,
\eea
with  two free parameters $\cal{N}$ and $\phi_0$,  turns to be enough for the  purpose of our fit.

The  function $D_{bg}$  should mimic the non-resonant background but also it  prevents the known Unitarity failure of otherwise no-unitary
Gounaris-Sakurai model, from which we have borrowed the peak functions ${\cal W}^{GS}_{\rho^i}$ (note $\Im$ stands for imaginary part). The constant phase approximation $\phi=\phi_c$ 
was taken to describe $\pi\pi$ phase shifts in a kindred 
fit model consideration \cite{BYKANA2016}. However herein, we have  found the inclusion of the second term in the Eq. (\ref{bgfaze})
 (this phase space function  with constant prefactor $1/3$ was chosen empirically)
represents  substantial improvement when minimizing the IE.

The function (\ref{mainfit}) then would represent $2+3N$ parametric fit for $N$ participating
  vector mesons, each of them is  associated with the single function ${\cal W}^{GS}_{\rho^i}$.
We found the three parameters characterizing the meson resonance would be highly correlated and we further reduce the number 
of free parameters by requiring the absolute value of the coupling $c$ is equal  for all $\rho $ meson excitations.
The entire fit thus has $7+2n$ parameters for $n$ excited states of $\rho$ meson, each of them have its own 
"width" $\Gamma$ and central mass $m$.

In order to get  fit of the $e^+e^-\to \pi\pi$ cross section we use the data collected by  CMD2 \cite{pipiCMD2005},
 SND \cite{pipiSND2006}  detectors as well as the ISR method extracted data by  BaBaR \cite{pipiBABAR2012}, KLOE \cite{pipiKLOE2005} 
 and BESS-III \cite{pipiBESSIII}. Within four wide and equally coupled resonances added to the admixture of $\rho(775)$ and $\omega(780)$ 
 mesons   we have achieved $\chi_I=1$ with the inflated error such that 
 $c_I=0.6 nb^{1/2}$ for $n_{dof}=548$.

Switching off the contribution from $\rho(1250)$   
 we would finish with much worse fit (large $\sigma_I$). More precisely, not allowing the meson with centered mass below $1400 MeV$ and using only three functions
 $\cal{W}_{GS}$ for excited $\rho$'s we get $\chi^2=2.2$ now for $n_{dof}=546$ (using  conventions established in preceding section). In this respect
the lightest resonance is confirmed according to \cite{rupp}. It has  a striking size of its width $\Gamma \simeq m$ , which is a  consequence of the presence of the 
function $D_{bg}$. In our case, three of four mesonic excitations have negative couplings and  only the third is positive. Furthermore, one should mention a not well pronounced but well known property of vector  resonances - most of them have negative couplings. We expect that the associated residua of  pole positions in the second Riemann sheet have their signs determined by couplings. From this it would  follow that using meaningful effective field theory for the accurate  description of resonances
one would need to accept negative kinetic terms.

We also compare with the popular fit based on the Gounaris-Sakurai (GS) model (used for instance by BaBaR collaboration \cite{pipiBABAR2012}). 
The GS fit  provides closer line to data when compared to  the analytic fit without inclusion 
of $\rho(1250)$, however the function   $ \chi^2=1.6$, which  is worse then our  analytical model (again, the error of the best fit was used). 
  Alternatively , one can say the inflated error of GS model is larger ($c_I=0.75$ for  our combined data).  Let us stress that our fit has
only 15 free parameters, which can be compared to GS model  (19 
free parameters). We did not purse further the GS model  since it violates the Unitarity.
\\
\begin{figure}[htb]
\centerline{\includegraphics[width=10.0cm]{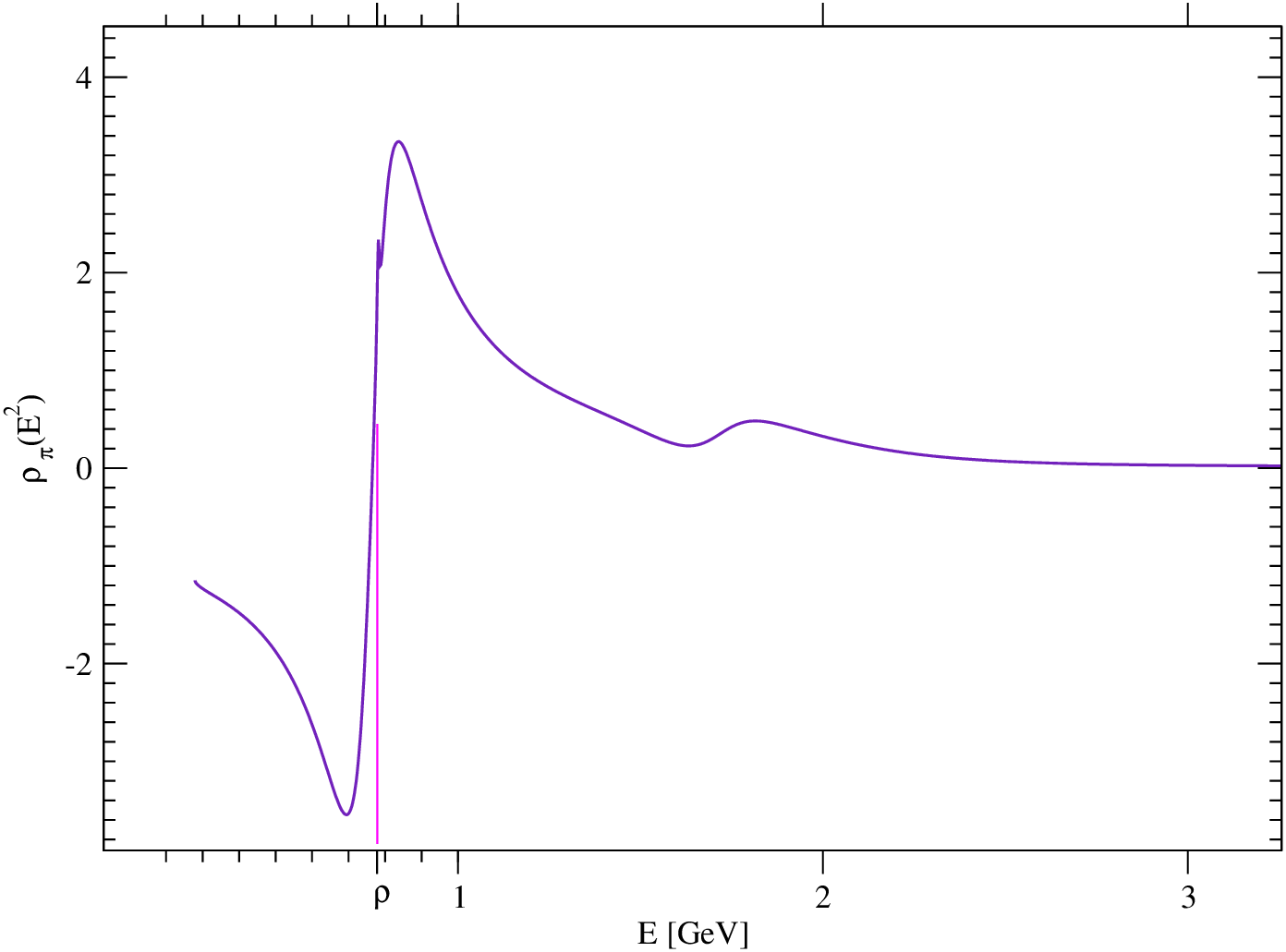}}
\caption{\label{nufik4}Spectral function of the pion electromagnetic form factor $F_{\pi}$. The peak for $\omega$ is visible slightly above $\rho$ meson mass, the later scale  is indicated by the pointer and letter $\rho$. A naively expected spectral peak of the  main $\rho$  resonance is  distorted  due to the presence of the background function 
$D_{bg}$. }
\end{figure}

Associated spectral function $\rho_F$  for the most likelihood (labeled AI) fit is shown in the Fig. \ref{nufik4} .
Contrary to the vacuum polarization spectral function, the function  $\rho_F$  is not positive definite.
 Requiring the positivity, we  would get resonances more localized ( see the Tab. for  small widths achieved),
 however the value of   $\chi^2\simeq 6$  excludes  this possibility as a reliable one.
 Interested reader can find  fitted values  in the table \ref{tabrho}, where also parameters for the GS fit are shown for purpose of completeness.
Although the meaning is lost, we still  use the word ``width'' for various $\Gamma$ parameters used in associated fits.

\begin{center} 
\begin{table}
\begin{tabular}{ |c|c|c|c|c|c|c|c| }
\hline
 fit                 & A I           &                 &             & A II        &(constr)       & G.S.       & model \\
\hline
name             &  m           &$\Gamma $&  c          &   m         & $\Gamma$ &   m         &  $\Gamma$ \\
\hline
$\rho(775) $   &  778.48   & 146.379     &  -          &   780.02  &  152.3        & 773.25    & 148.6        \\
$\omega$       &  781.684 &  8.4500      &    0.001514 &   781.69  &  8.53          &  780.4    &  9.609       \\
$\rho(2)$        & 1202.0    &   1270.0    & -0.0666  &  1299.0   &  150.0        &    ---      &  ---           \\
 $\rho(3)$       & 1694.0    &  704.00     & -0.0666  &  1428.6   &  230.0        & 1356.5   &  252.2       \\
 $\rho(4)$       & 1735.0    &  291.7       &  0.0666  &   1742     &  134.0        & 1668     & 124.0         \\
 $\rho(5)$       & 2200.0    &  1242.0     &  -0.0666 &   2217     &  150.0        &  2217    & 320.0         \\
 \hline
 $\chi^2$ /n.o.p.       & 1.0         &/ 15 par.     &               &    6         & / 18 par.      &  1.6       & / 19 par.\\  
\hline
\end{tabular}
\caption{\label{tabrho}List of parameters  for various fits as described in the text. AI stands for the most likelihood fit based on spectral representation, AII stands for the 
same but with with constrained positivity and G.S. stands for 19 parametric GS model. $\chi^2$ and number of free parameters are shown, $\chi^2$ functions are  calculated with IE of AI for all fits for purpose of comparison. In order to avoid confusion a simple numbering is used to name the excited  $\rho$ mesons.  }
\end{table}
\end{center}

All combined data   and the most likelihood fit is  shown in  Fig. \ref{nufik} (left panel) some details, e.g. chosen  errors  are shown in right panel of  the Fig. \ref{figure5}.  
 The constant phase in the Eq. (\ref{bgfaze}) was fitted such that  $\phi_o=-1.1303 rad$, the overall normalization factor of the spectral function $\rho_F$ 
 was found ${\cal{N}}^{-1}= 0.85560 $. Couplings for two additional fits are different, they are complex for GS fit and not displayed for brevity. 
 Further details of GS fit are displayed in the Appendix.

To conclude this Section we argue that  our 15 parametric  analytical fit based on the spectral representation for the pionic form factor excellently describes the world combined data on two pions productions. Adding  a new meson entity does not  lead to substantial reduction of the 
inflated error. In implies,  already right number of  degrees freedom was achieved  and  the size of the IE  
is limited by  property of the data rather then by quality of our  fit.

\section{$\Pi_h$ determination}

In the previous section we have found numerical fits for the most important contribution to HVP.
 The inflated error $\sigma_I(s)$  was determined by using the global $\chi^2$
 criterion for each model fit separately. Form this point of view the 15 parameters spectral representation fit and 19 parameters GS model fit provided  comparable good results, however their different analytical forms can lead to the local changes 
which are not completely clear from the global characteristic of $\chi^2$ fits.
 In order to see these effects we evaluate the muon production cross section
and compare with precise measurement at phi meson peak as well as we compare electromagnetic running charge as has been measured 
by KLOE collaboration more recently.

 Likewise in  approaches \cite{polar2,polar3,DAVIER2011}, 
the method of obtaining of HVP herein is based on the integration of the Eq. (\ref{muf}), where in our case the  smooth function $\sigma_{fit}$ is  used 
for this purpose. In order to achieve our desired  comparison one needs to add all remaining important  contributions to $\sigma_h$.
Following the similar  routine as in the calculation of $\sigma_{\pi\pi}$ contribution, we have found fits with acceptably small continuous IE for the remaining exclusive channels: $K^{+}K^{-}$, $K_LK_S$ and $\pi\pi\pi$
 as well as sub-dominant  $\eta\gamma$ and $\pi\gamma$ production cross sections have been  included.
 With the precision achieved the  final states with four pions has been  included as well, while we have neglected $KK\pi$ and the other states with higher multiplicity then 3.
 Neglected  cross sections are flat, very small and their quantitative effect 
was estimated by comparison with off resonances HP functions calculated in \cite{polar2,polar3}, where these channels were recently included. 
Further, following the standard routine, the effect of well established vector  charmonia and bottomonia  has been included into $\sigma_h$  
via  their BW forms with PDG  averaged values.

All necessary fits with sources of experimental data are described in the appendices, each separately devoted to the individual process as named above.
For this purpose we profit from  a number  of accurate and compatible  measurements as well as from a number of published  
  well established and known interpolating fits for all necessary 
 exclusive channels. For each case we however perform our own $\chi^2$  and determine IE. Fro the later, a  using of formula  (\ref{maj}) fully accepts the last experiments,
 eg.  CMD-3, BESS-III, KLOE, as well as most  SND,CMD-2 and BABAR measurements, while we have    discard the data from  old experiments (CMD,DM,NA7,OLYA,TOF).
 For each individual exclusive cross section we use just one fit, with the exception of the   cross section $\sigma_{\pi\pi}$ discussed in preceding Section. 
All codes and required data inputs are collected at author's web page \cite{www}.

The second necessary ingredients to maintain precise comparison one needs  to solve (\ref{muf}) with relatively high numerical precision.
The so called Hollinde trick was used to perform principal value integration (for the meaning see for instance  Eq. (C.2) in the paper \cite{SAULI2003}). 
 The integral equation (\ref{muf}) was solved iteratively within a combined integration methods (Simpson and Gaussian) 
at large numerical grid. In case of the use of the analytical fit for $\sigma_{\pi\pi}$ we have used the identical grid to evaluate the fit and  the HVP.
In this way we get rid of additional systematical numerical error due to the interpolation and have achieved numerical precision required for a meaningful comparison of tiny effects made by the  use of different fits for $\sigma_{\pi\pi} $ cross section.

\subsection{$\alpha_{QED}$ at $\phi$ meson energy}

\begin{figure}[htb]
\centerline{\includegraphics[width=8.0cm]{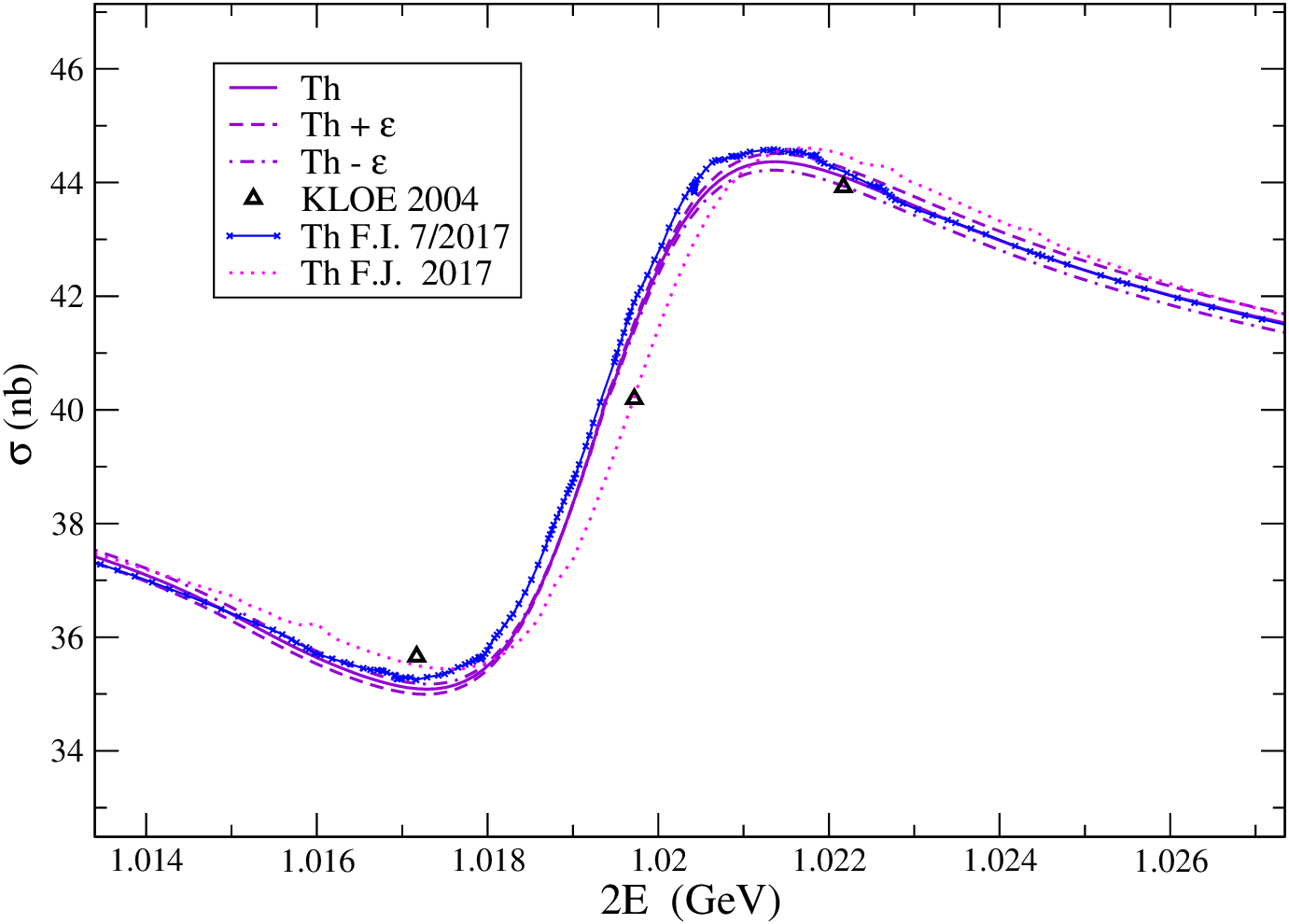}}
\caption{\label{figure1}Muon pair cross section, comparison between theory and experiment as described in text.}
\end{figure}

Let us start with the comparison of calculations for the muon pair productions at $\phi$ meson energy.
HVP obtained by method described in previous Section and by other methods  in  \cite{polar2,polar3} is used to calculate $\mu\mu$ cross section and compared 
to the experiment in the Fig. \ref{figure1}. 
The line labeled by "Th F.J." or "Th F.I." represents   cross section calculated with the use of HP in \cite{polar2}  and in \cite{polar3} 
respectively.  
\\
-
\\

Observed $\phi$-meson  kick in $\mu\mu$ spectrum is  reproduced by 
the Standard Theory dispersion relation and it is shown in the Fig.\ref{figure1}. the  error band (dashed and dot-dashed lines) is added.

 The differences among various theoretical results have systematical origin due to the  
different strategy used for  determination of HVP at $\phi$ meson invariant mass. 
Remind that the total error was governed by systematical error due to the luminosity and detection uncertainties
 ($\delta_{syst}=1.2 \% $) all  "theoretical" lines agree with experiments within experimental errors.
The  error band (dashed and dot-dashed lines) is added into the figure for comparison as well.
We do not show the error band for the calculation based on AI fit.  Instead for purpose of figure  we have used $c_I=0.8$ to determine the error band,
  recalling  the fit based on the spectral representation for $\sigma_{\pi\pi}$  generates the same result at $\phi$ meson energy as the fit based entirely on $GS $ model.
 The figure is thus identical to those in   \cite{SAHVP1,SAHVP2}.  The KLOE data points are represented by triangles, noting that the  standard statistical deviations roughly correspond with the size of the triangle.  Slightly different situation appears at lower energy as it is discussed in the next Section.

\section{$\alpha_{QED}$ at KLOE2}

\begin{figure}[ht]
\centerline{{\includegraphics[width=7.60cm]{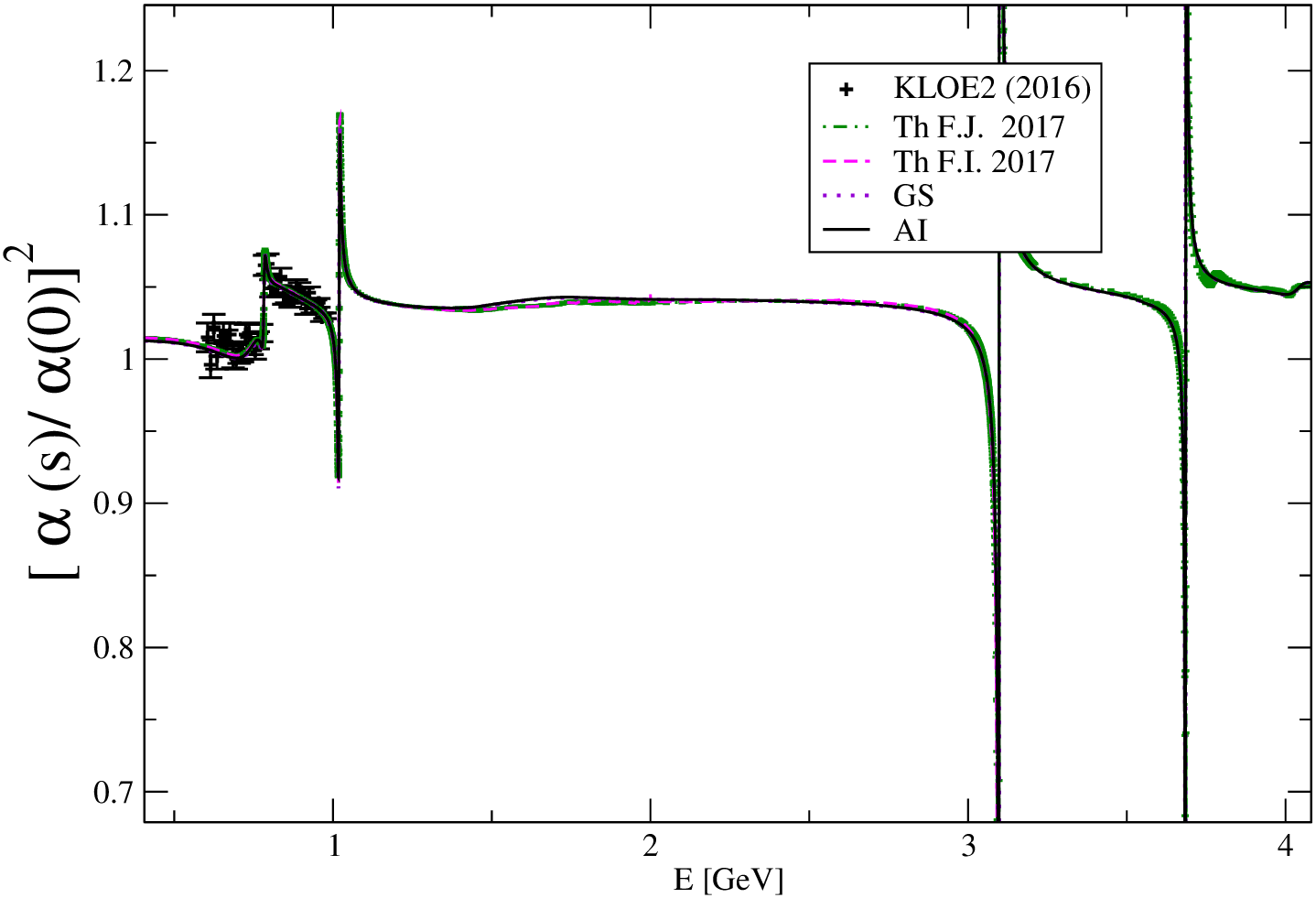}}{\mbox{-}\mbox{-}}
{\includegraphics[width=7.60cm]{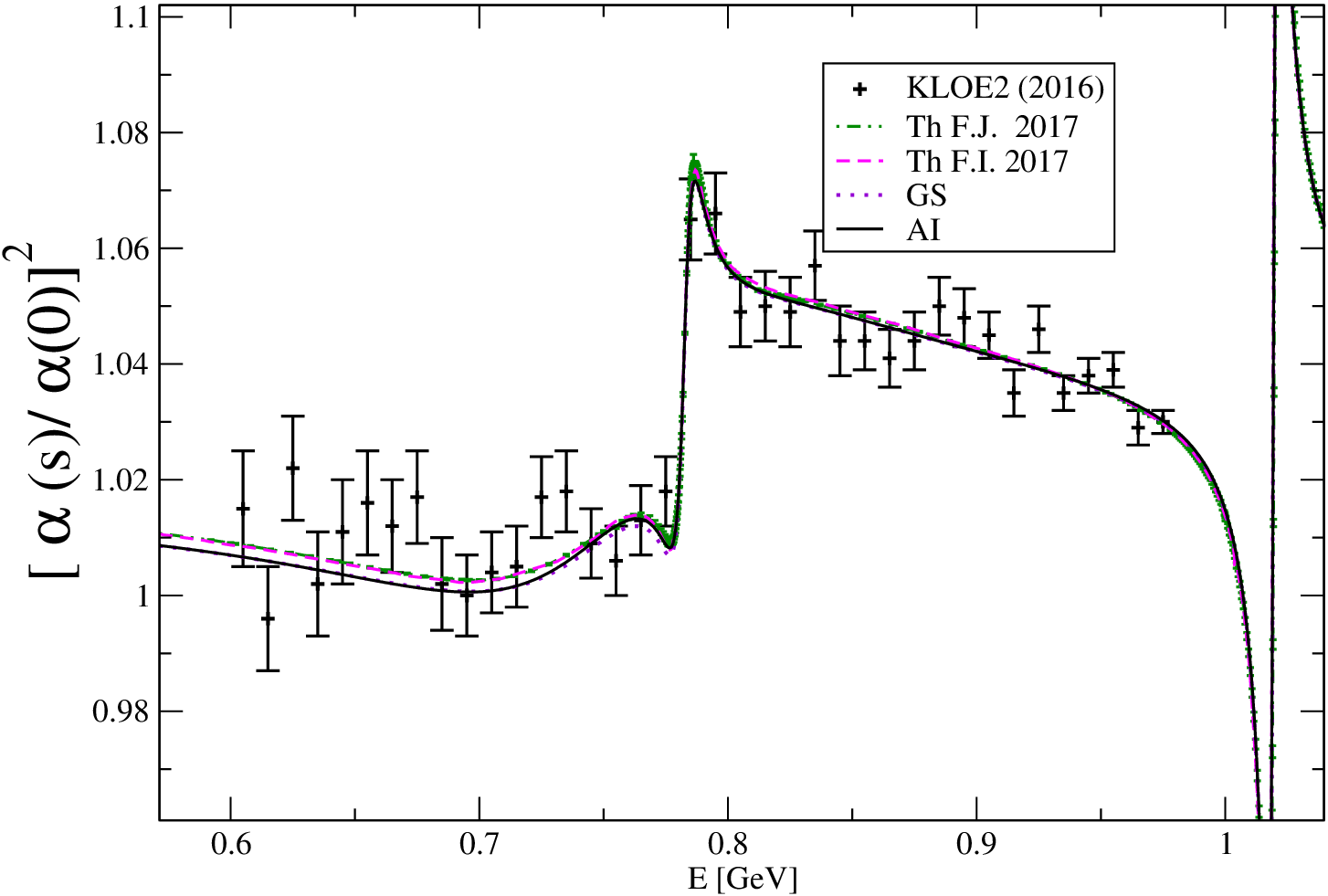}}}
\caption{Square of the fine structure constant. Lines labeled by F.I. , F.J  stand for HP extractions made by  \cite{polar3} or  \cite{polar2} respectively. 
Global view and comparison with KLOE2.}
\label{celkove}
\end{figure}

\begin{figure}[ht]
\centerline{{\includegraphics[width=7.80cm]{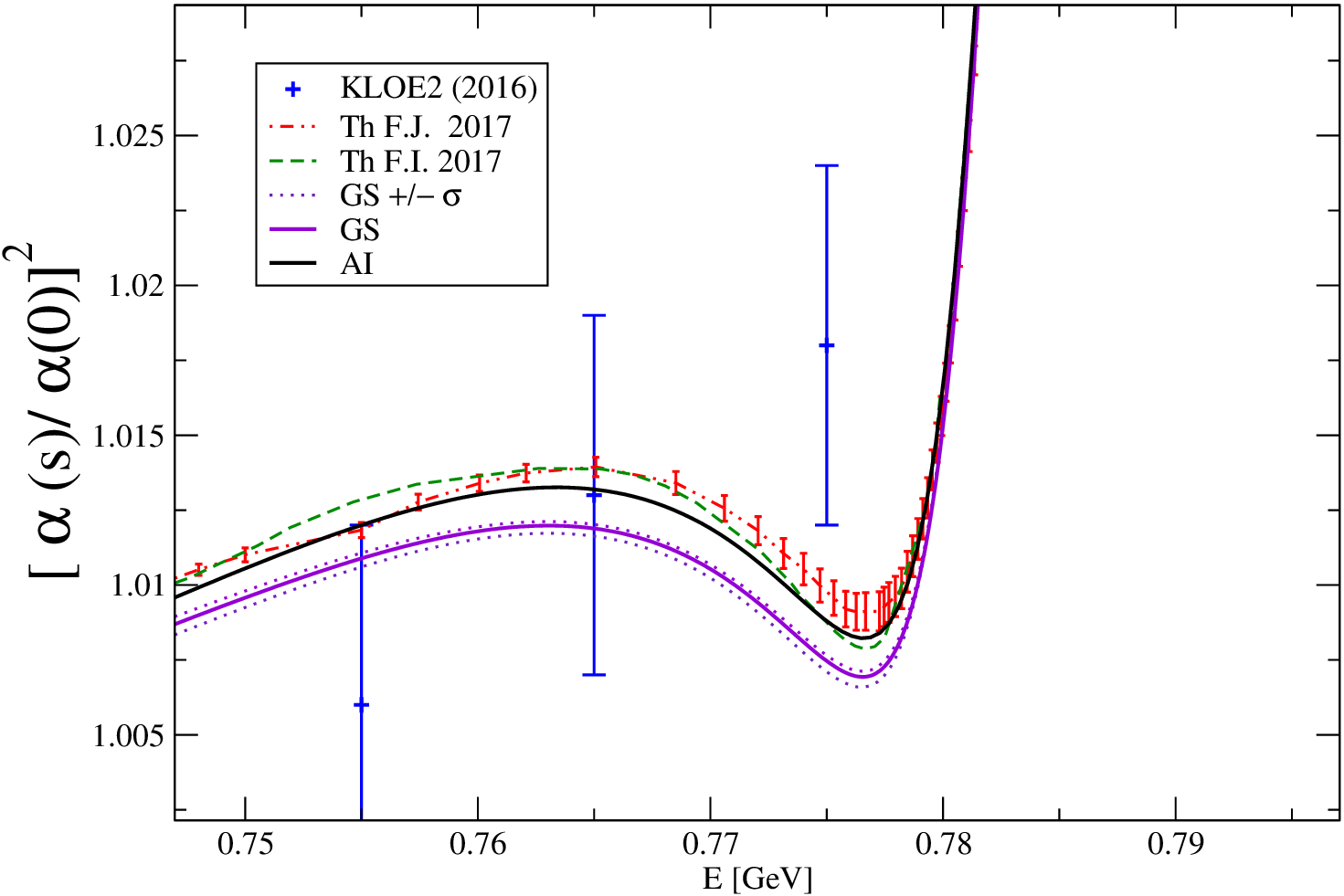}}{\mbox{-}\mbox{-}}
{\includegraphics[width=8.0cm]{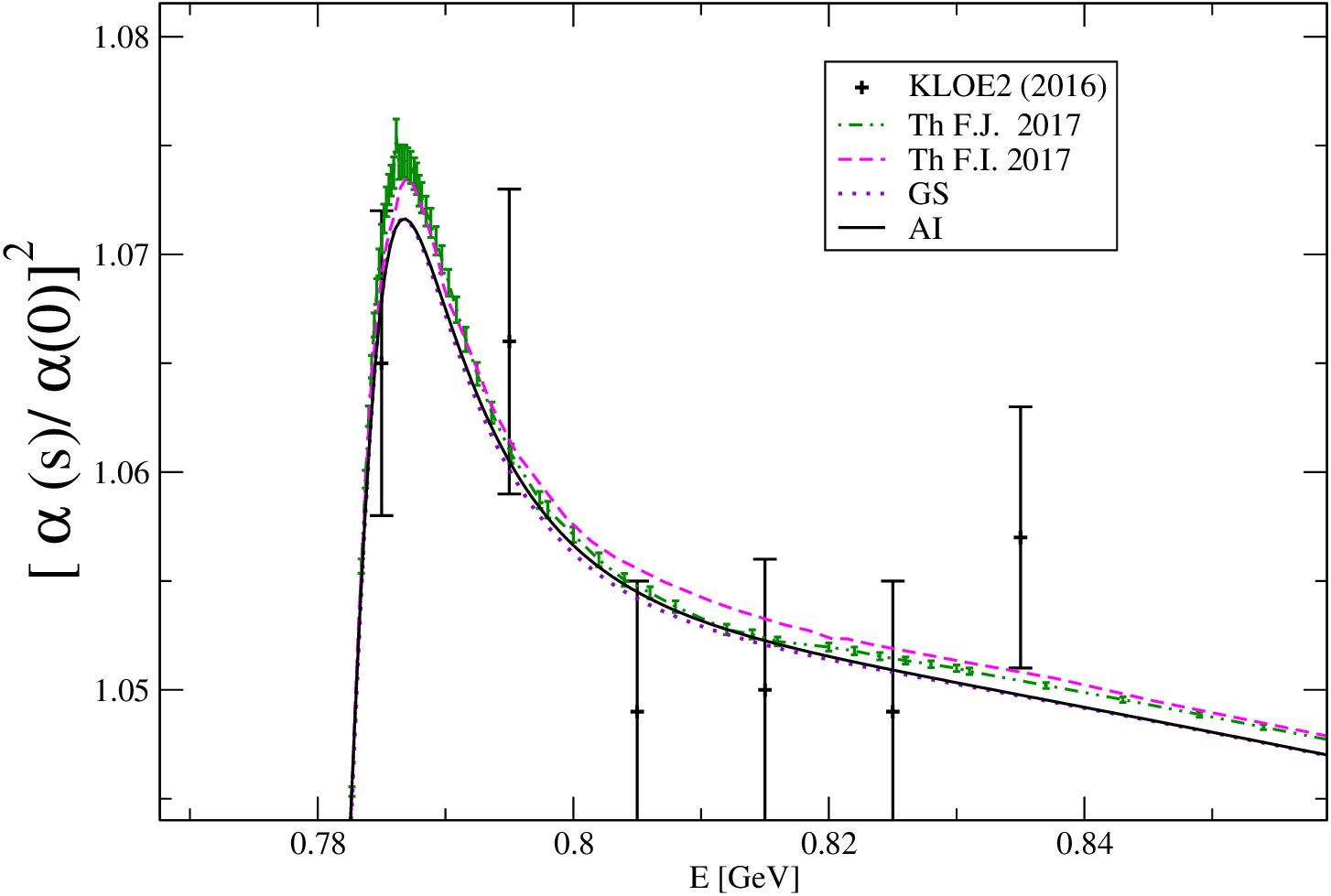}}}
\caption{\label{dethaily}Square of the fine structure constant as in the previous figure. Details at the shoulders of $\rho/\omega$ peak.\\}
\end{figure}

Very recently, the running of the electromagnetic coupling has been  measured via radiative return method  by KLOE2 collaboration \cite{KLOEdva}.
Thanks to achieved precision, it is  for the first time, when  the "kick" effect was seen in the $\rho /$$\omega$ energy region. 
All calculated  results for QED  running coupling are compared in four figures, more specifically  for the large region of $s$ in figures \ref{celkove} as well as  details for KLOE2  accessible region are shown for the left shoulder (left figure \ref{dethaily}) and the right shoulder  (right figure \ref{dethaily}). 
All determined  HP functions provide almost identical picture of the running coupling for KLOE2 energy region and there are very small
 difference between the result obtained here and in  \cite{polar2,polar3}.
A tiny error bands are not shown as they are about magnitude smaller then the standard experimental deviation.

In comparison to \cite{polar2,polar3}  we have got a smaller polarization at very near of the muonic threshold. The origin of this 
remains unclear to the author. Looking elsewhere in momentum axis it should not  originate in neglected high multiplicity final hadronic states
(interested reader can find   the portion of  individual hadronic contributions   for instance in \cite{SAHVP1,SAHVP2}).

{\mbox{}}
\\
\\
{\mbox{}}
\\

\section{Conclusion}

The strategy of evaluation of HVP based on the use  of fits for exclusive hadroproduction in $e^+e^-$ annihilation  was suggested and
 the evaluation was performed in practice.  We advocate for the use of special form for the inflated error which mix  statistical and systematic errors of several experiments and  is particularly suited for the analyzes of  combined data on $\sigma_h$ bellow $3GeV$.
In addition to some popular phenomenological fits, the dispersion relation for the pion
  electromagnetic form factor was found to be excellently working here. Within a reduced number of free parameters describing the form factor,
it provides best fit for combined data from the threshold $2 m_{\pi}$  to 2.5 GeV. 
However it turns that some of  identified resonances, including also reintroduced excited meson $\rho(1250)$, are largely affected  
by presence of  the background as well as  their mutual interference is enhanced and plays the role.
Nevertheless, the entire spectral function $\rho_{\pi}$  is considerably simple and can be used to evaluate  other physical quantities 
(e.g. anomalous magnetic moments of leptons).

The aforementioned spectral fit together with the other considered fits for $\pi\pi$ (hadronic) production cross sections have been then utilized to calculate of $\sigma_{\mu\mu}$ cross section and QED running charge at the timelike region of momenta and compared with 2004 KLOE  $\sigma_{\mu\mu}$ as well ass with 2016 KLOE2
measurement  for the electromagnetic running charge. In both cases the ``theoretical'' error due to the determined inflated error of $\sigma_h$ 
was shown to be several times smaller then the total experimental errors. Obtained results have been compared with those where hadronic vacuum polarization was extracted by other methods. Based on  the mutual comparison of different fits of $\sigma_h$,
 which have the similar global quality quantified by  $\chi^2$, allowed us to show that (difficult to estimate) systematical error due to the methodology  used herein is  reasonable small.

Regarding our best resulting fit for the process $\sigma_{\pi\pi}$, the  masses of $\rho(n)$ relatively  
 agree with the others (\cite{rupp}), whilst  widths are much larger in case of two first excited states.
 The reason is that there is a  relatively admixture with  the background functions $D_{bg}$, 
 which enhance the mutual interference among resonances as well.
 Obviously,  the overlap of resonances is such enormous, that the interpretation of 
 such contributions as radially excited mesons can be  a severe simplification of more complicated nonperturbative terms.
 To this point let us mention the observation \cite{sauli2020} , where it was shown that the quark dynamical mass function  
 is not a smooth function of momentum variable at the timelike region and 
no simple quantum mechanical model, which uses a constant quark mass approximation, 
can be therefore reliable  trusted. Actually, the running  of masses at the timelike region  can cause reordering or mutual washout of resonances and $\rho(2)$ and $\rho(3)$ vectors are not distinguishable in principle. Due to the same reasoning  the mass of $m_K^*$ does not need to be heavier then $m_{\rho(2)}$  if one needs to avoid very large meson  width for any reason. In a honest approach, a wide resonances can be regarded as an effective description of more non-perturbative contributions mixtured at once.

\section{Acknowledgments}
The author is grateful to P. Byd\v{z}ovsk\'y for stimulating discussions and thanks  to prof. F. Jagerlehner for his help and  advice to run the code \cite{polar2} and 
 to G. Venanzoni 
for  invitation to FCCP 2017, which was partially motivating and useful for final comparisons with the  data of other groups.

\appendix

\section{Details on analytical fit}

In this appendix we review the list of  functions, which have been used in our  constructions.

Especially important is the form of Gounaris-Sakurai dressed vector meson propagator, which reads   
\be  \label{GSprop}
{\cal W}^{GS}=\frac{m^2+d(m)\Gamma/m}{M^2(s)-s-i m\Gamma(s,m,\Gamma)} \, \, ,
\ee
\be
M^2(s)=m^2\left[1+\frac{\Gamma k^2(s)}{k^3(m)}(h(s)-h(m^2))+\frac{\Gamma h^{'}(m^2)}{k(m)}(m^2-s)\right] \, \, ,
\ee
\bea
\Gamma(s,m,\Gamma)&=&\Gamma\frac{m}{\sqrt{s}}\left[\frac{L_2(s,m_{\pi})}{L_2(m^2,m_{\pi})}\right]^3 \, \, ,
\nn \\
L_2(s,m_{\pi})&=&\sqrt{s-4m^2_{\pi}} \, \, , \label{eldva}
\eea
where we have defined following auxiliary functions:
\be
h(s)=\frac{\beta(s)}{2}\ln\left(\frac{\sqrt{s}+2k(s)}{2m_{\pi}}\right) \, \, ,
\ee
\be
 h^{'}(m^2)=\frac{2 m_{\pi}^2 h(m)}{m^4\beta(m)}
+\frac{2 m_{\pi}^2}{\pi m^4 \beta(m)}
+\frac{\beta(m)}{2\pi m^2} \, \, ,
\ee
\be
d(m)=\frac{4m_{\pi}^2}{m^2\beta^3(m)}(3h(m)-2/\pi)+\frac{1}{\pi\beta(m)} \, ,
\ee
with the usual shorthand notation:
\be
\beta(s)=\frac{L_2(s,m_{\pi})}{\sqrt{s}} 
\, \, ,\, \, 
k(s)=\frac{L_2(s,m_{\pi})}{2} \,\, ,
\ee
 used for the velocity and the  two pion Lorentz invariant phase space factor.

 The Breight-Wigner function for the narrow $\omega$ meson was taken in the form:
\be
{\cal W}_{\omega}=\frac{m^2_{\omega}}{m^2_{\omega}-s-im_{\omega}\Gamma_{\omega}} \, .
\ee

\section{Gounaris-Sakurai fit of $ F_{\pi}$}

The G-S fit for combined data reads
\be \label{fitrho}
F_{GS}(s)=\frac{{\cal W}^{GS}_{\rho}(s,m_{\rho},\Gamma_{\rho}){\cal D}_{\rho}(s)\frac{1+c_{\om}{\cal W}_{\om}(s,m_{\om})}{1+c_{\om}}
+\sum_{i}c_{\rho^{i}}{\cal W}^{GS}_{\rho^i}(s,m_{\rho^i},\Gamma_{\rho^i})}{1+c_{\rho^{1}}+c_{\rho^{2}}+c_{\rho^{3}}} \, \, .
\ee
where the entering functions ${\cal W}^{GS}$ and ${\cal W}_{\omega}$ are  identical to those defined  in the previous appendix.
However here they are taken with  complex prefactors: 
$c_V=|c_V|e^{i\phi_V}$. In addition we have found
advantageous to deform $\rho/\omega$ peak by the introduction of auxiliary function ${\cal D}_{\rho}$, which was chosen such that  ${\cal D}_{\rho}=1$ above $\rho$ meson mass and 
\be
{\cal D}_{\rho}(s)=x+(1-x)\left[\frac{1-4m_{\pi}^2/s}{1-4m_{\pi}^2/m_{\rho}^2}\right]^{1/2} \, ,
\ee
bellow the value  $s=m_{\rho}^2$. The parameter $x$ was fitted providing the value $x=0.37202$. 

The index i runs over the resonances labeled  as $\rho(1500)$, 2 for$ \rho(1800)$ and $i=3$ stands for BaBar resonance corresponding with the peak at  $2200 MeV$  in the 
paper \cite{SAHVP2}, while they are relabeled for purpose of this paper as $\rho(3), \rho(4)$ and $\rho(5)$ respectively.

\section{Fit for $\pi^+\pi^-\pi^0$ channel}

\begin{figure}[htb]
\centerline{\includegraphics[width=10.5cm]{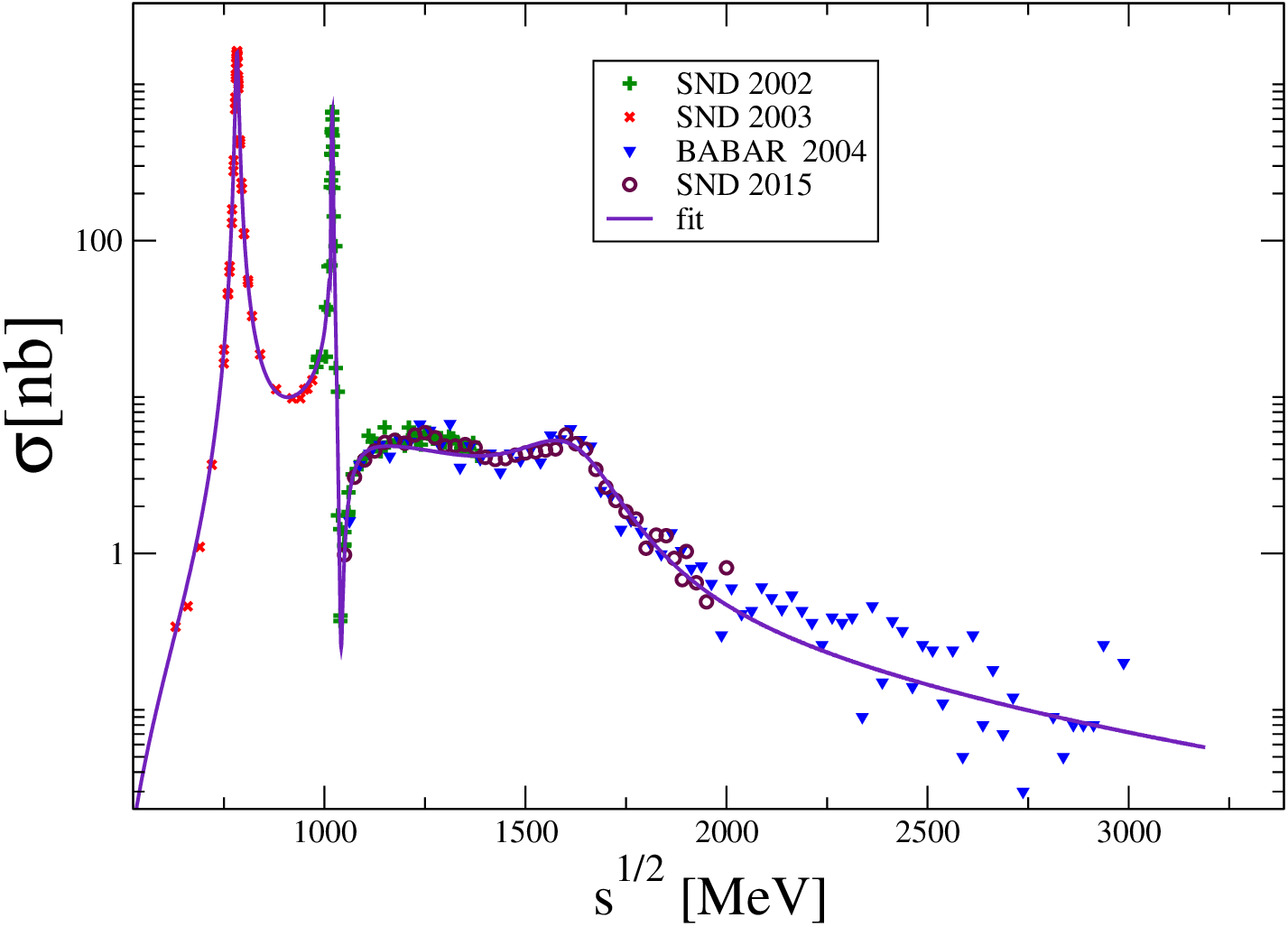}}
\caption{Measured $\sigma_{3\pi}$ and the fit, global view.}
\label{figure8}
\end{figure}
\begin{figure}[htb]
\centerline{\includegraphics[width=9.0cm]{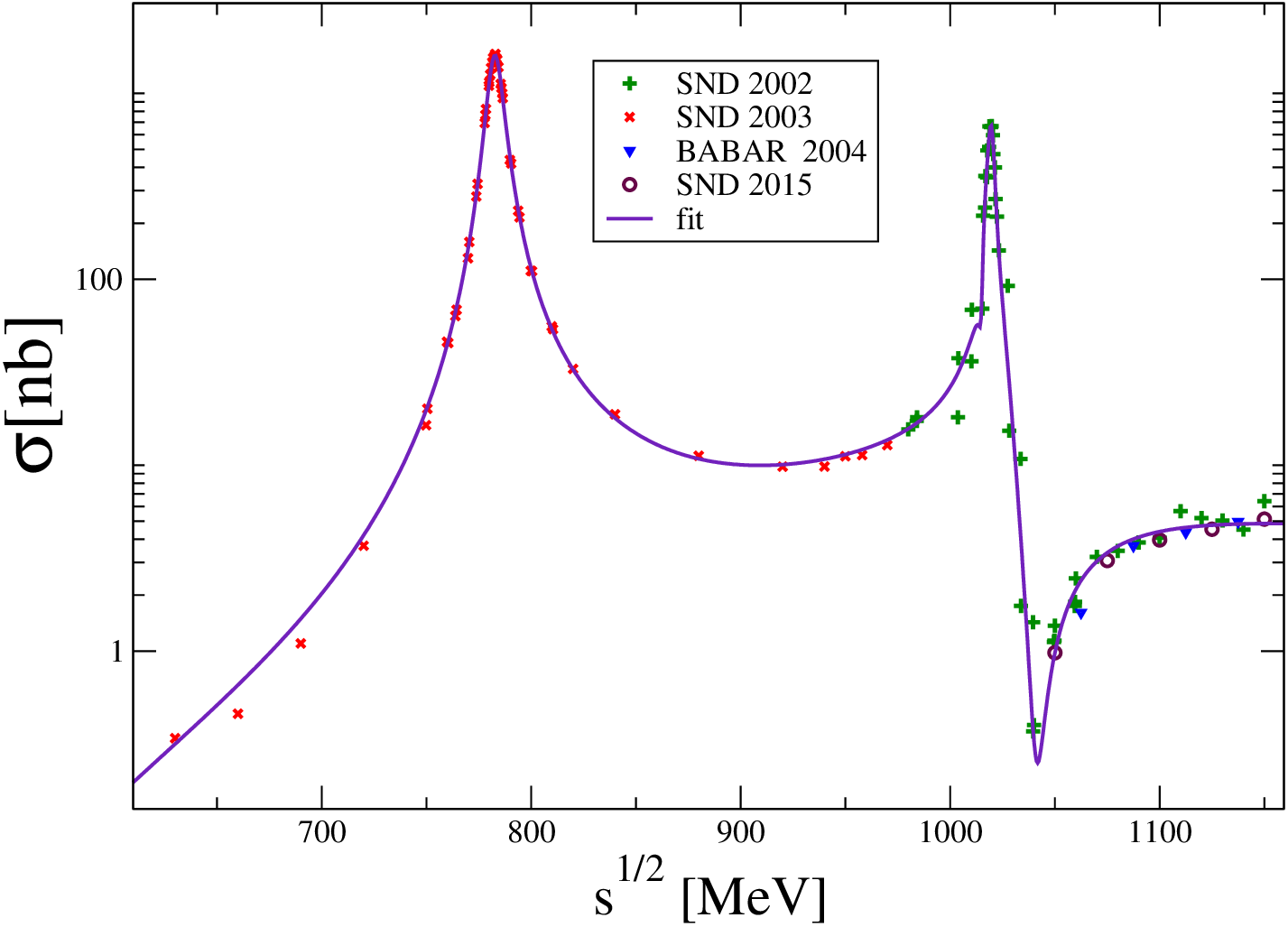}}
\caption{Selected data for  $\sigma_{3\pi}$ and the fit, view on the peaks.}
\label{figure9}
\end{figure}

 
Complicated by the shape, $ e^+e^-\to 3\pi $ total cross section consists from two dominant peaks of the narrow omega and phi vector resonance. The first peak can be  represented by  an almost perfect BW function, while the second one is crudely deformed as the $\phi$ meson peak turns abruptly down and makes the fitting  more complicated.
Happily a  sum of complexified  and slightly deformed BW functions provide  very good auxiliary function for making a fit out of the data even without taking of full correct three pions phase space and without the use of any  "background function".  However these are the data itself which does not allow to minimize $\chi^2$ with the same  error function as in the previous case and the IEF is taken slightly larger by enlarging the  coefficients in
 Eq. (\ref{maj}) such that  $c_L=1 nb^{1/2}$ and   $c_s=1 $ . In this exceptional  case we get minimized $\chi^2$ slightly larger then one: $\chi^2=1.23$ with the resulting curve and the data shown in the Fig. \ref{figure8} and in the Fig. \ref{figure9} in detail. Note also  here, that we get $\chi^2=0.95$ if we cut the  data above $1.05 GeV$, having thus the region of $\omega$ and $\phi$ mesons under a better control. 

Like in  the previous case, only BW functions with masses higher 
then the threshold are used. We do not exploit  VMD  idea of  rho meson as an intermediator, wherein virtual decay 
$\omega \rightarrow \pi\rho \rightarrow 3\pi$ would require "dressed"  rho meson propagator and numerical integration over the  three body phase space would be needed. This would inevitably  causes a drastic  grow of the time  of the minimization procedure (from days to unacceptable years).  Actually, we are not improving a given VMD model but we are looking for the  smallest $\chi^2$ instead, for which purpose the use of proposed auxiliary functions is more suited in practice. 

Our simplified fit therefore reads:
\bea
\sigma_{3\pi}(s)&=&\frac{12\pi}{s^3}L_3(s)
\left|\sum_{V=\omega,\phi}
{\cal W}_V {\cal D}_V+
\sum_{V=1,2,3}
{\cal W}_V\right|^2 \, \, ,
\nn \\
{\cal W}_V&=&\frac{m_V\Gamma_{V}(s) e^{i\phi_V}}{m_V^2-s-im_V \Gamma_V(s)}
\sqrt{\frac{m_V B_V}{L_3(m_V^2)^3}} \, \, ,
\label{brejt}
\eea

where for all $V$ the auxiliary functions now read
\bea
\Gamma_V(s)&=&\Gamma_V\frac{m_V L_3^3(s)}{s^{1/2}L_3^3(m_V^2)} \, ,
\nn \\
L_3(s)&=&\sqrt{s-9m_{\pi}^2} \, ,
\eea
and where  the function which further deform omega meson peak reads
\bea
{\cal D}_{\omega}(s)&=&\left[1+c_1\frac{s-m_{\omega}^2}{2}
+c_2\sqrt{s-m_{\omega}^2}\right]^{-1}
\Theta(s-m_{\omega}^2)+
\nn \\              
&+&\left[1+c_3\frac{(s-m_{\omega}^2)}{2}
+c_4\sqrt{m_{\omega}^2-s}\right]^{-1} \Theta(m_{\omega}^2-s) \, \, ,
\eea
with fitted constants $c_1=0.105321 {\mbox GeV}^{-2}$, $c_2=-0.0598 {\mbox GeV}^{-1}$, 
 $c_3=-0.254221 {\mbox GeV}^{-2}$, $c_4=0.055406 {\mbox GeV}^{-1}$.

Whilst the function which deforms BW shape of $\phi$ meson is chosen 
different from $\omega$ and reads
\be \label{kotatko}
{\cal D}_{\phi}(s)=1+{\cal W}_{L}(s)\Theta(m^2_{\phi}-s)+{\cal W}_{R}(s)\Theta(-m^2_{\phi}+s) \, ,
\ee
where in Eq. (\ref{kotatko})  BW functions  (\ref{brejt}) are taken. Stress here, that 
these functions serve to deform the shape  of the  left and the right shoulder of $\phi$ meson resonance and  they appear in the product with $\phi$ meson BW and should not be confused with a conventional meson. All fitted numbers  are listed in Tab. (\ref{omegatab}). In the function $L_3$ the value $m_{\pi}=139.57018 {\mbox MeV}$ is taken, ignoring  the difference between charged and neutral pion mass. The cross section is taken from $\sqrt{s}=3m_{\pi}$, bellow it is zero.

\begin{center}
\begin{table}
\begin{tabular}{ |c|c|c|c|c|}
\hline
   --    &   mass/MeV &   width/MeV &  B &  $ z $ \\
$\omega$  &782.6141  & 8.7031  &  $3.0885\,\, 10^{-5} $  &    0 \\
L       &1015.461 &  2.813   & 0.167           &      134.1 \\
R      & 1025.487  & 12.301  &   0.121          &     184.35   \\      
$\phi $   &1019.704  & 4.067   &  $5.0807 \,\, 10^{-5}$  &    163.62   \\                  
$V_1 $   & 1086.425  & 252.0   &  $  8.9  \,\,  10^{-7} $ &  124.36 \\
$V_2 $    &1219.280  & 569.0   &   $ 2.97 \,\,   10^{-7} $  &  78.123 \\
$V_3 $   & 1636.32  &  278.0    &  $ 1.52  \,\,  10^{-6} $  & 174.62 \\
\hline
\end{tabular}
\caption{\label{omegatab} Parameters of the   cross section fit for $\sigma_{K^+K^-}$ as described in the text.}
\end{table}
\end{center}

In usual VMD's the parameter $B$ stands for the  product of branch ratios $Br(V\to ee)Br(V\to 3\pi)$. Here the value for $\omega$ meson significantly differs from the BaBaR measurement ($Br(\omega\to ee)Br(\omega\to 3\pi)_{\mbox BaBaR}=6.7\,\,  10^{-5}$) since   the fit is different as well. There are other differences, whether  stemming from our different formula for the fit is not obvious. The last resonance agrees with the meson conventionally labeled as $\omega^{''}$ (see \cite{3piSND2015}), noting also that $\phi^{'}$ observed at the same energy in other process (see the next Section) should be there as well.
On the other side, there is no good evidence  for super-wide ($\Gamma\simeq 900 MeV$) SND/BaBaR established meson. This, over all overlapping resonance, conventionally labeled  $\omega^{'}$, with quoted mass $\simeq 1470\pm 50$ by SND2015,  is preferably 
replaced by two BW functions with much lower masses and different complex couplings. 
 Of course, recalling the meaning and purpose of our fit, which uses complex phases and avoids a use of  correct 3-body phase space, does not allow to make a strong statement about the vector meson content of $\sigma_{3\pi}$ cross section. 

-
\\

\section{Fit for cross section of the process $e^{+}e^{-} \to K^{+}K^{-}$}

Very important, since the most dominant exclusive process at $\phi$ meson peak energy, has been measured not only 
on the peak, but thanks to the ISR method  also fairly above: up to the total energy $E=8 {\mbox GeV}$.
Fine selected data are chosen from several last measurements, e.g. the most precise data \cite{charcmd2} are fully taken into account, we  have also used selection survived  (off peak) data as obtained by SND \cite{chargeSND1,chargeSND2},  and from the energy 1350 MeV till 5 GeV we exploit the  BaBaR data \cite{chargebababar1,chargebababar2}, providing total $N_{d.o.f}=142$ for $K^+K^-$ cross section.  Remind two important notes here for completeness. Firstly, the  $J/\Psi$ and $\Psi'$ peaks were subtracted by BaBaR collaboration and we add them separately. Secondly, keeping a certain amount of threshold BaBaR data is possible, in a way one still keeps $\chi^2<1$ without changing fit. Here we simply  preferred to keep $\chi^2$ lower, with $N_{d.o.f.}$ smaller for future purposes.  Fit for the  charged K meson pair production cross section reads
\be
\sigma_{K^+K^-}(s)=|A|^2 s^{-5/2}f_{c}\left[\frac{L_2(s,m_K)}{L_2(m_{\phi}^2,m_K)}\right]^3
\ee
with the function $f_c$ in used is defined as 
\bea
f_{c}&=&\frac{1+\alpha\pi(1+v(s)^2)/2v(s)}{1+\alpha\pi(1+v(m_{\phi}^2)^2)/2v(m_{\phi}^2)}      \,\, ,
\nn \\ 
v(s)&=&L_2(s,m_K)/\sqrt{s} \, \, ,
\eea
where $m_K$ is the mass of charged Kaon, $L_2$ is defined earlier in (\ref{eldva}),  and the amplitude is given by the sum of $BW$ functions:
\be \label{common}
A=\sum_V e^{i z_V} {\cal W}_V(s)=\sum_V e^{i z_V}\frac{\sqrt{c\,  m_V\,  B_V\,  \Gamma_V }}       
         {m^2_V-s-i\Gamma_V\sqrt{s}}\, \, ,
\ee
where $c=4.5465 \,\,  10^{-5}$, and where all BW functions are common for narrow as well as for wide resonances.

\begin{figure}[ht]
\centerline{\includegraphics[width=12.0cm]{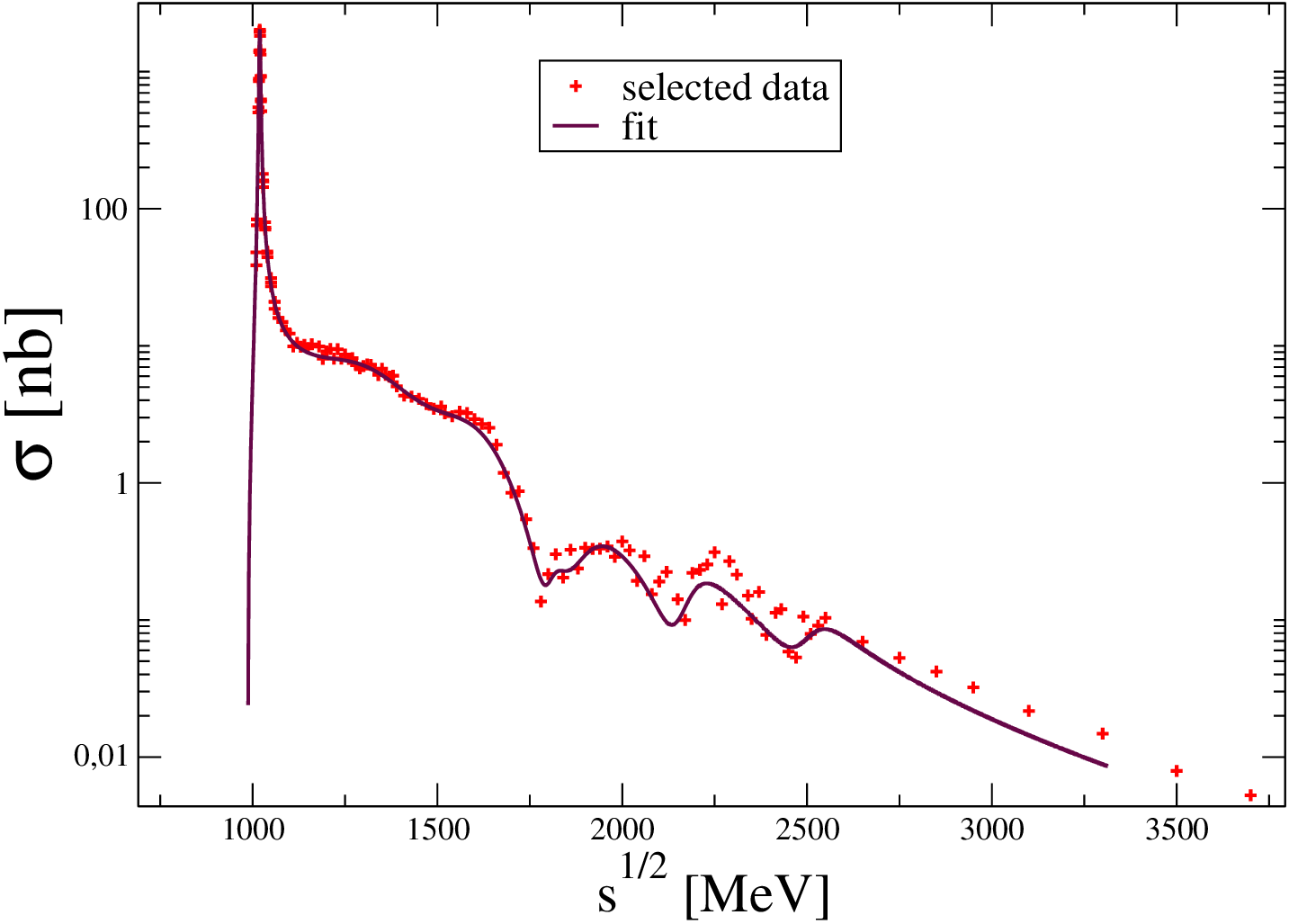}}
\caption{Selected data for  $\sigma_{K^+ K^-}$ and the fit, $\phi$ peak view.}
\label{figure7}
\end{figure}

The sum in  (\ref{common}) runs over the BW functions, noting that the four of lowest five can be identified with usual
i.e. more or less established radial excitation of the $\phi$ and $\omega$ . Their names  are quoted in bracket in the first column of the Tab. \ref{ficka}. The one unlabeled there  has a small coupling to the leptons and do not need to be necessarily  related with  conventional meson, however it helps to accommodate  the shape of fit to the cross section data.
Up to the ground state meson, we are not strongly pointing a given BW structure with a given meson name, since  due to the interference effect the parameters are strongly correlated. In general, it is hard to label overlapping resonances, noting trivially  that the observed pattern above $\phi$ meson arises from the admixture of the light flavor quark-antiquark components, however what is   flavor content of a   single broad BW peak is not obvious, at least when comparing to  electrically neutral ground state vectors: $\phi,\omega$ and $\rho$ mesons. Selected data and the  fit with $\chi^2=0.56$ (obtained with IEF) are illustrated at Fig. \ref{figure7}.

{\mbox{}}
\\
                
\begin{center}
\begin{table}
\begin{tabular}{ |c|c|c|c|c|}    
\hline      
  label(content)                  & mass/MeV &   width/MeV &  B &  $ z $ \\         
  $\phi$                          & 1019.2469 & 4.1358   &  0.5377    &  0   \\
    $V_1(\omega^{'})$             & 1337.1    &372.83    &  0.00890   & 180.71\\
   $ V_2(\phi^{'}\omega^{''})$    & 1624.8    & 307.406  &  0.0061    & 222.53 \\  
      -                           & 1813.033  & 80.01    &  0.0003535 &  92.6\\
     $V_3(\phi^{''})  $           & 1892.36   & 262.0    &  0.00252   & 104.91\\
      $V_4$                       &  2178     &  157.7   &  0.000175  & 100.2\\
      $V_5$                       &  2510     &  160.1   &  0.0000364 & 134.2\\
\hline
\end{tabular}
\caption{ \label{ficka} Numbers for charged kaons channel}
\end{table}
\end{center}

\section{Fit for the neutral kaons channel $e^{+}e^{-} \to K^{0}_SK^{0}_L$}
-
\\
\begin{figure}[htb]
\centerline{\includegraphics[width=9.0cm]{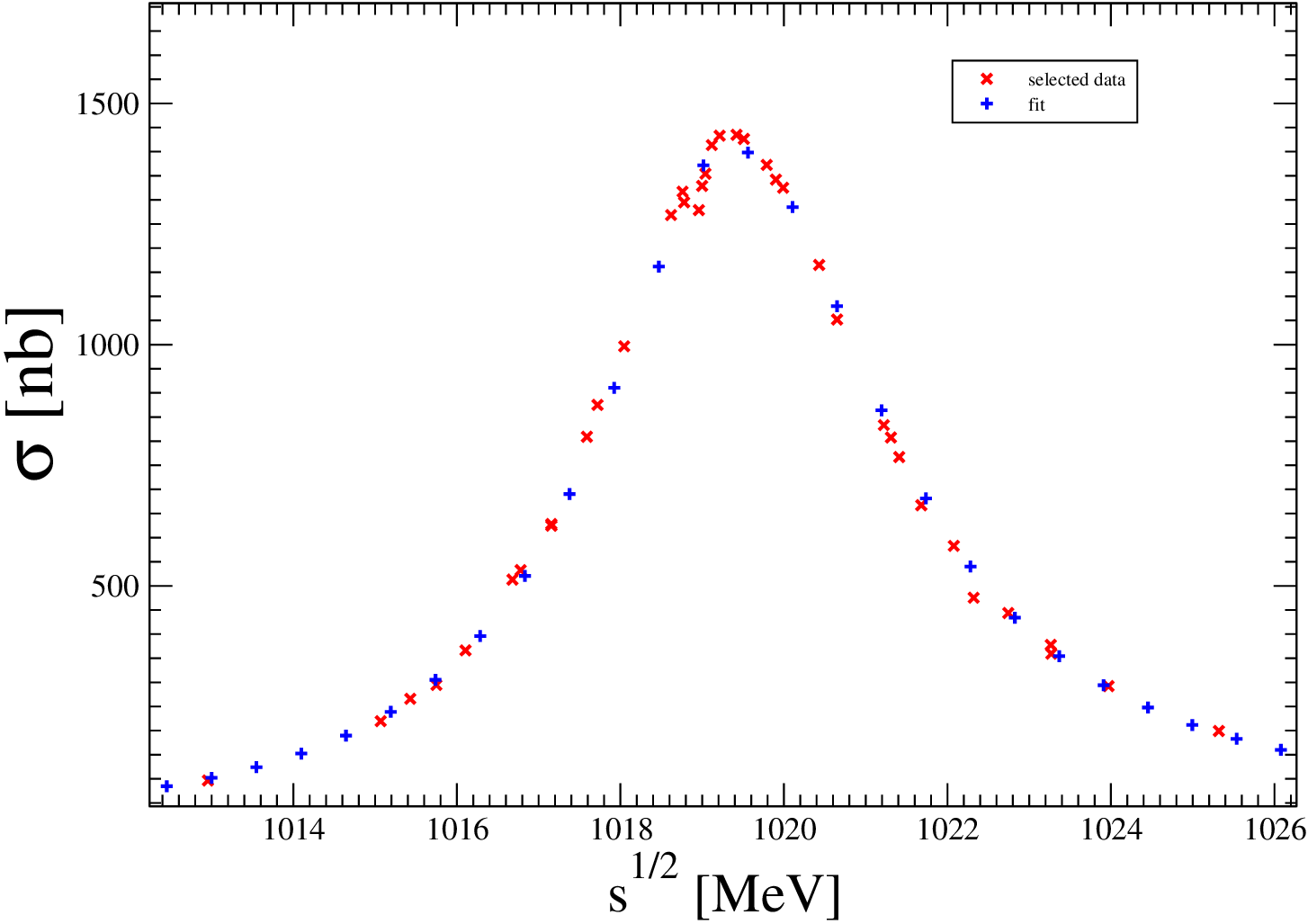}}
\caption{Measured $\sigma_{K_L K_S}$ and the fit around the peak position.}
\label{figure3}
\end{figure}

To fit the process  $\sigma(e^{+}e^{-} \to K^{0}_SK^{0}_L)$ we have used the data collected by Novosibirsk SND,CMD-2 collaborations and very newly by  CMD-3 group \cite{SNDKK2001,CMD2KK2004,CMD3KK2016} as well as the BaBaR data \cite{BABARKK2014} were exploited above $\phi$ meson peak.
For this purpose  the similar formula as for the process $e^{+}e^{-}-->K^{+}K^{-}$ is used, however 
in addition to that, we have introduced  the deformation function into the cross section (i.e.  there is a change
$|A|^2 \rightarrow d(s)|A|^2$), where the function $d$ is represented by  the following step functions
\bea
d(s)&=&\left[1+\frac{d_1 (s-m_{\phi}^2)}{2}
+d_2\sqrt{s-m_{\phi}^2}\right]^{-1}
\Theta(s-m_{\phi}^2)
\nn \\              
&+&\left[1-\frac{d_3(s-m_{\phi}^2)}{2}
-d_4\sqrt{m_{\phi}^2-s}\right]^{-1} \Theta(m_{\phi}^2-s)
\eea
and  four fitted numbers  within four(five) digit accuracy read:
$ d_1=1.3204 {\mbox GeV}^{-2}, d_2=0.8615 {\mbox GeV}^{-1}, d_3=12.371 {\mbox GeV}^{-2}, 
d_4=1.291 {\mbox GeV}^{-1}$.

The amplitude is made out solely from BW functions, wherein we have found that three vector mesons are enough. However, due to the flatness of the cross section, it is advantageous to distinguish the  narrow $\phi$ meson and  wide resonances. The appropriate BW functions   read:
\bea
{\cal W}_{\phi}(s)&=&\frac{m_{\phi}^3\sqrt{12\pi\Gamma_{\phi}B_{\phi}/m_{\phi}}}{m^2_{\phi}-s-i \Gamma_{\phi} \frac{s}{m_{\phi}}}\, \, ,
\nn \\   
{\cal W}_{V}(s)&=& e^{i z_V} \frac{m_V \Gamma(s,m_V) \sqrt{B_V m_V/L_3(m_V^2,m_{\pi})}}{m^2-s-i m_V \Gamma(s,m_V)} \, \, ,
\nn \\
\Gamma(s,m_V)&=&\frac{m_V\Gamma_V}{\sqrt{s}}\left[\frac{L_3(s,m_{\pi})}{L_3(m_V^2,m_{\pi})}\right]^3\, \, , 
\eea 
where we have used   $L_3$ function for   the running width for
resonances $V_{1}\simeq \phi^{,}$ and $ V_2\simeq \phi^{,,}$. 
Relevant numbers are listed in the Tab. \ref{neutral} :

\begin{center} 
\begin{table}
\begin{tabular}{ |c|c|c|c|c|}
\hline          
            & mass/MeV &   width/MeV &  B &  $ z $ \\         
$\phi$      &1019.3886 &4.2612  & $0.4235\, 10^{-6}$ & -- \\
$V_1$  &1670.320  & 198.0  &$ 0.24975\,  10^{-6}  $ &177.5 \\
$V_2$ &2066.985  & 248.5  & $0.3072\,  10^{-6}   $ &147.8 \\
\hline
\end{tabular}
\caption{Numbers for neutral kaons channel} \label{neutral}
\end{table}
\end{center}

\begin{figure}[htb]
\centerline{\includegraphics[width=10.0cm]{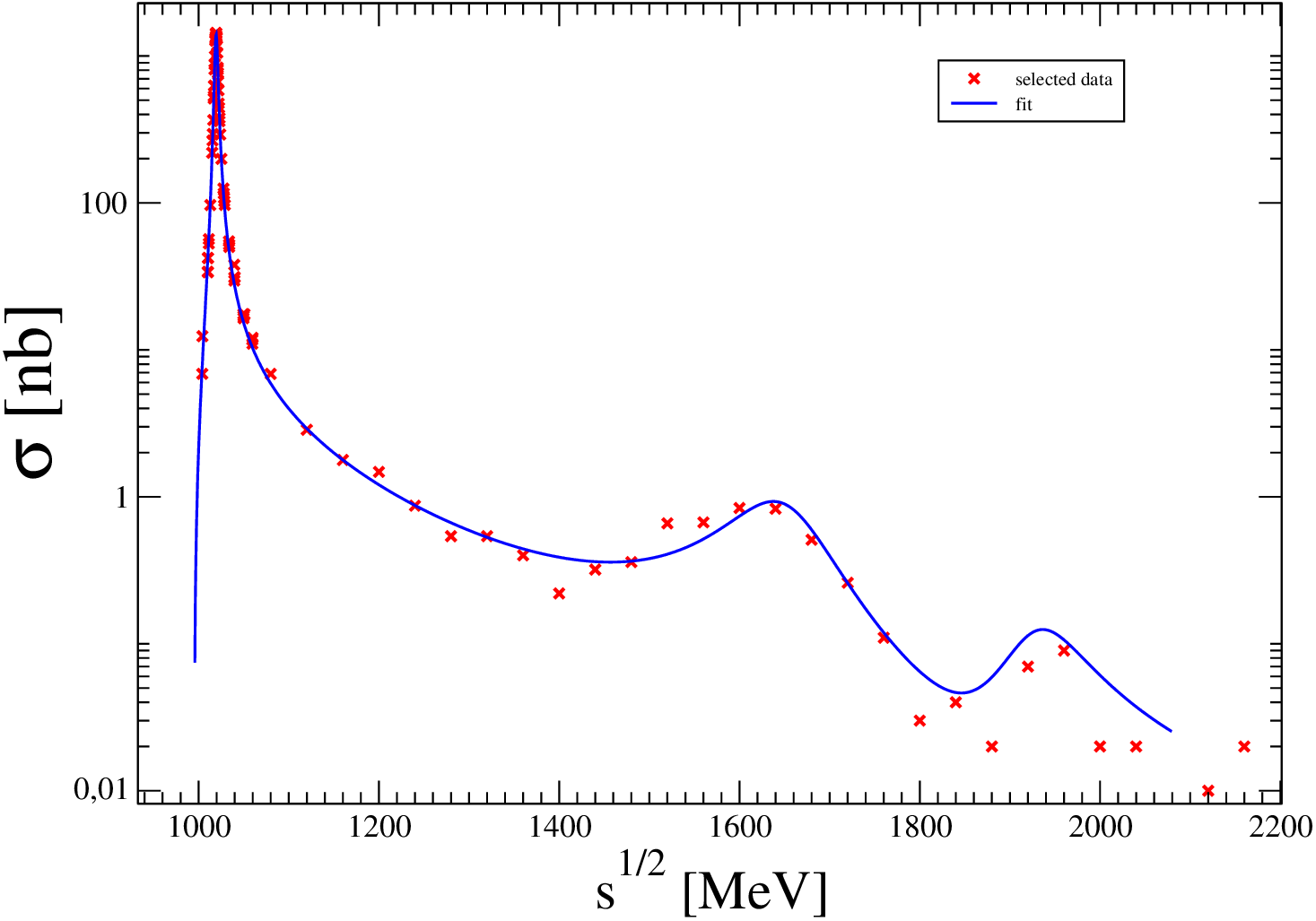}}
\caption{Measured $\sigma_{K_L K_S}$ and the fit, global view.}\label{figure4}
\end{figure}

To get the fit the data with $N_{d.o.f}=93$ has been used. The results are  sketched in  figures \ref{figure3} and \ref{figure4}.
\\
%%%%%%%%%%%%%%%%%%%%%%%%%%%%%%%%%%%%%%%%%%%%%%%%%%%%%%%%%%%%%%%%%%%%%%%%%%%%%%%%%%%%%%%%%%%%%%%%%%%%%%%%%%%%%

\section{Fit for the $\eta\gamma$ and $\pi\gamma$ cross sections}
-
\\
-
\begin{figure}[htb]
\centerline{\includegraphics[width=9.0cm]{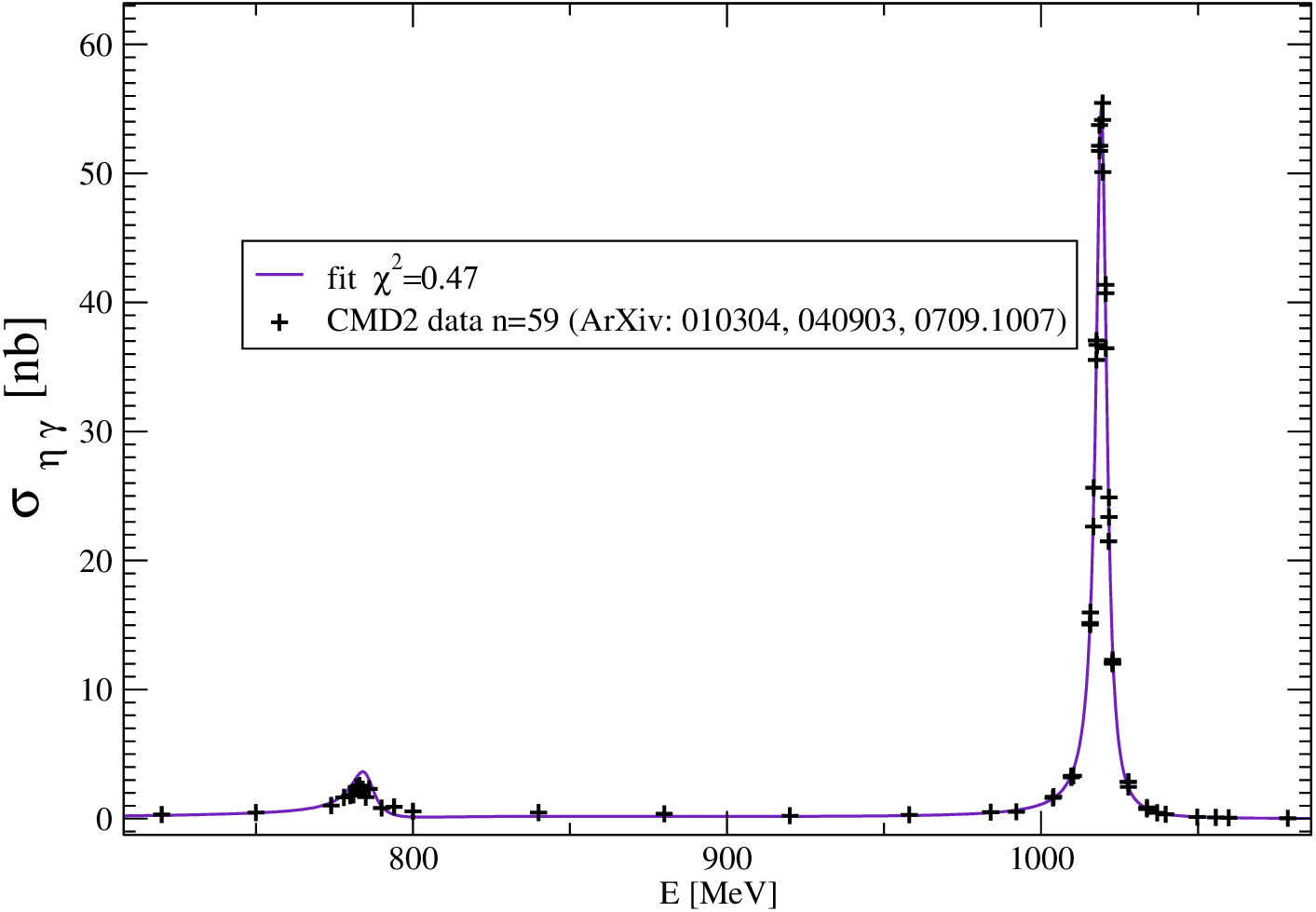}}
\caption{Measured $\sigma_{\eta\gamma}$ and its fit.}\label{figure2}
\end{figure}

\begin{figure}[htb]
\centerline{\includegraphics[width=9.0cm]{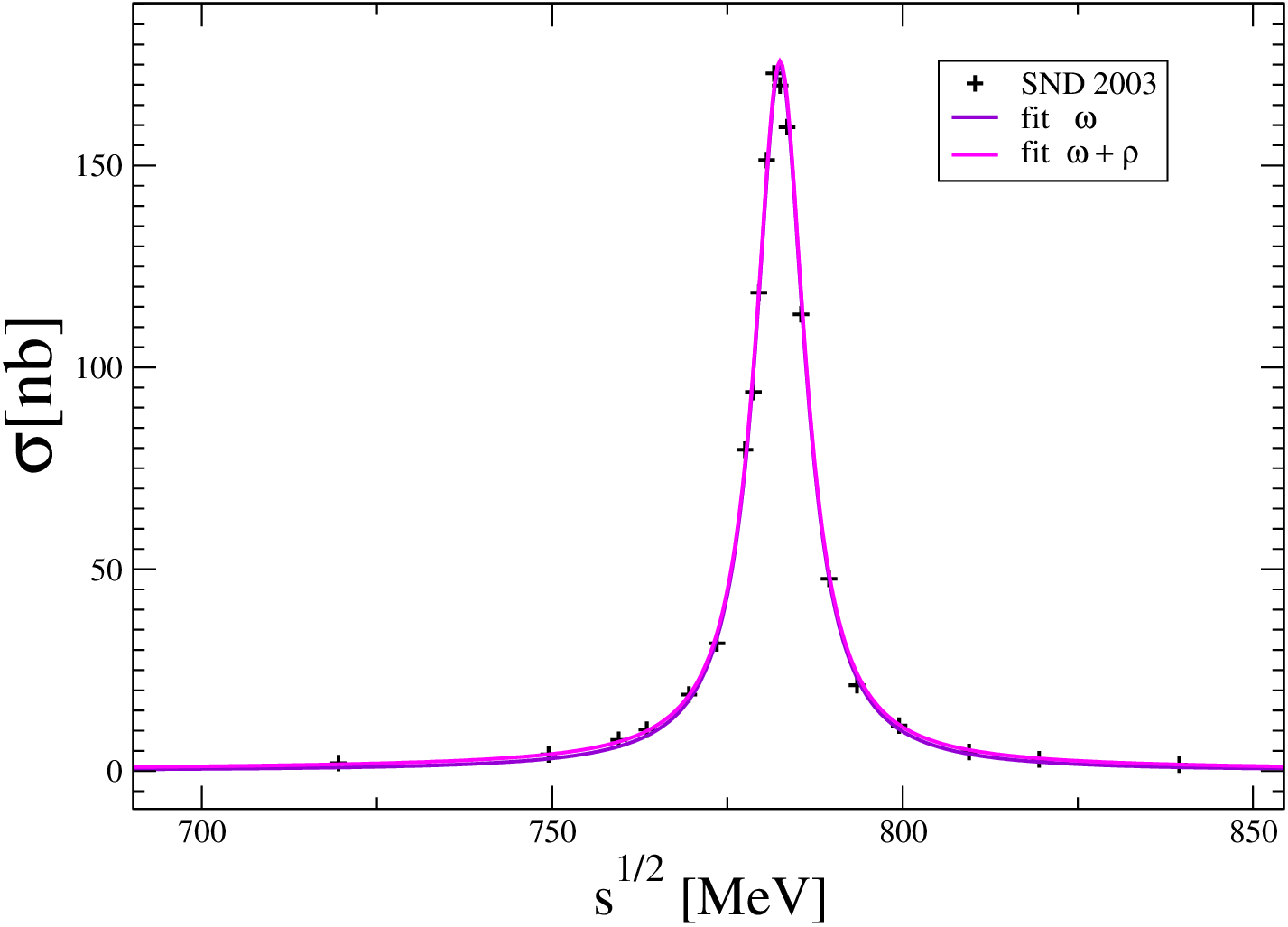}}
\caption{Measured $\sigma_{\pi\gamma}$ and two fits as described in the text are shown.} \label{pigam}
\end{figure}

\begin{figure}[htb]
\centerline{\includegraphics[width=9.0cm]{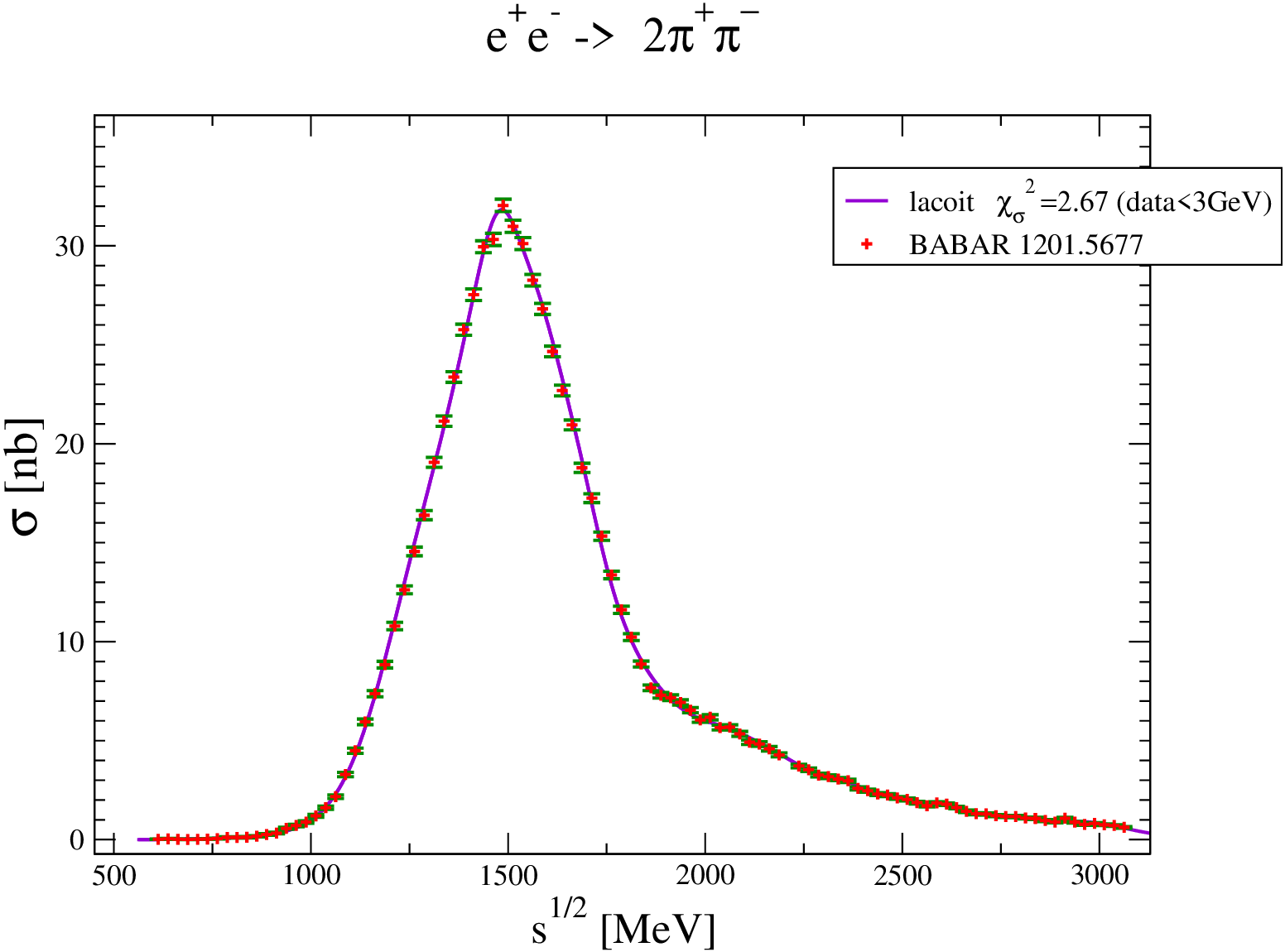}}
\caption{Measured $\sigma_{4\pi}$ and the fit.}\label{posle}  
\end{figure}

The fit of the  cross section of the process  $e^{+}e^{-} \to \pi^0 \gamma$ is based on simple 
admixture of $\rho$ and $\omega$ BW functions with constant parameters. It reads
\be
\sigma_{\pi\gamma}(s)=K(1-m_{\pi}^2/s)^{3/2}
\left[\frac{m_{\omega}^4}{(m_{\omega}^2-s)^2+\Gamma_{\omega}^2 m_{\omega}^2}
+2.1\frac{m_{\rho}^4}{(m_{\rho}^2-s)^2+\Gamma_{\rho}^2 m_{\rho}^2}\right]
\ee
with $K= 64/(3 m_{\pi}^2) 10^{-9}$ and the remaining parameters are $m_{\omega}=782.5 MeV, \Gamma_{\omega}=8.63 MeV;
m_{\rho}=775.02 MeV, \Gamma_{\rho}=149.59 MeV$. We neglect the interference term in this case.
The resulting fit is shown in the  Fig. \ref{pigam} where also the fit with only  $\omega$ meson is shown for interesting comparison (numbers not shown). 
The better fit, the one with  inclusion of $\rho$ meson, is used for the calculation of hadronic $\Pi_h$.

For the cross section $\sigma_{\eta\gamma}$  we have used the data with $N_{d.o.f}=59 $  measured by CMD-2/SND detectors and published in the period 2001-2014 
\cite{etagam1,etagam2,etagam3}, where also the data for $\sigma_{\pi\gamma}$ has been obtained.

The parameterization of the cross section was chosen such that
\be
\sigma_{\eta,\gamma}(s)=\frac{1}{s}\left[{\cal W}_{\omega}(s,m_{\omega})+
e^{iz_{\phi}}{\cal W}_{\phi}(s,m_{\phi}){\cal D }(s)\right]^2 \, \, ,
\ee
where the choice (\ref{common}) for BW was made.  Only $\omega $ and $\phi$ mesons are considered and heavier mesons
are ignored as the cross section for $\eta,\gamma$ production  is fairly small \cite{etagam3} at higher energies.
The deformation is considered for the $\phi$ meson case, while the phase space factor
is effectively absorbed into the fit.

The fitted values of  BW parameters are $m_{\omega}=785.9 MeV, \Gamma_{\omega}=9.06 MeV;
m_{\phi}=1019.415 MeV, \Gamma_{\rho}=4.0306 MeV$ and the deformation function is
chosen as
\bea
{\cal D}(s)&=&\left[1+\frac{24.49 GeV^{-2}(s-m^2)}{2}
+0.1326 GeV^{-1}\sqrt{s-m_{\phi}^2}\right]^{-1}
\Theta(s-m_{\phi}^2)+
\nn \\              
&+&\left[1+\frac{-4.62 GeV^{-2}(s-m2)}{2}
-3.0 GeV^{-1}\sqrt{m_{\phi}^2-s}\right]^{-1} \Theta(m_{\phi}^2-s)\, \, .
\eea
The phase was fitted such that $z_{\phi}=20.65^o$. The fitted function and the data are shown in the Fig. \ref{figure2}.

\section{Fit for the $4\pi$ cross section}

To fit the cross section  $\sigma(e^{+}e^{-} \to 4\pi)$ we used the data collected  by BaBaR collaboration \cite{ctyripi}. 
Working with single measurement  the experimental statistical error can be used to find a fit, in this case $\chi^2=2.7$
was achieved, providing  safe constrain $\chi^2<1$ when IEF is used instead.
As a fitting function we have chosen the product of two BW's series with a suitable phase factor. It simulates $2\rho$ peak with other six heavy excitation in combination included as well.
Further, the shape is modified by the function made out of the sum of unit and  the eight properly centered  Gaussian functions. 
Like in previous case, the  fit serves for numeric and  should not confused with any  "effective theory" calculation and it should be regarded as such. 
We do not write down the full details of the fit, noting only here, it involves 28 parameters related with BW functions and their mixing, and 24 parameters
related with Gaussian functions (for interested reader, the code can be sent via email on  request of the author). The number of parameters are not used to reduce n.d.f. (number of data points).       
The fitted function and the data are shown in the Fig. \ref{posle}.

%%%%%%%%%%%%%%%%%%%%%%%%%%%%%%%%%%%%%%%%%%%%%%%%%%%%%%%%%%%%%%%%%%%%%%%

%
\end{document}